\documentclass{article}

\usepackage{arxiv}

\usepackage[utf8]{inputenc} 
\usepackage[T1]{fontenc}    
\usepackage{hyperref}       
\usepackage{url}            
\usepackage{booktabs}       
\usepackage{amsfonts}       
\usepackage{nicefrac}       
\usepackage{microtype}      
\usepackage{lipsum}
\usepackage{float}
\usepackage{graphicx}
\usepackage{subcaption}
\usepackage{tikz}

\usetikzlibrary{trees}
\hypersetup{
    colorlinks=true,
    linkcolor=blue,
    filecolor=magenta,      
    urlcolor=cyan,
}

\title{Deep learning for understanding multilabel imbalanced Chest X-ray datasets}

\author{
 Helena Liz \\
  Dept. Computer Sciences, Universidad \\
  Rey Juan Carlos\\
  Computer Systems Department, \\
  Universidad Polit\'{e}cnica de Madrid, Spain\\
  \texttt{helena.liz@urjc.es} \\
 \And
Javier Huertas-Tato \\
Computer Systems Department, \\
Universidad Polit\'{e}cnica de Madrid, Spain\\
\texttt{javier.huertas.tato@upm.es} \\
   \And
Manuel Sánchez-Montañés \\
Computer Science Department, \\ Universidad  Aut\'{o}noma de Madrid, Spain \\
  \texttt{manuel.smontanes@uam.es} \\
   \And
 Javier Del Ser \\
 TECNALIA\\Basque Research \& Technology Alliance (BRTA), Spain\\
  \texttt{javier.delser@tecnalia.com} \\
     \And
 David Camacho \\
  Computer Systems Department, \\
  Universidad Polit\'{e}cnica de Madrid, Spain\\
  \texttt{david.camacho@upm.es} \\
}

\begin{document}
\maketitle

\begin{abstract}

Over the last few years, convolutional neural networks (CNNs) have dominated the field of computer vision thanks to their ability to extract features and their outstanding performance in classification problems, for example in the automatic analysis of X-rays. Unfortunately, these neural networks are considered black-box algorithms, i.e. it is impossible to understand how the algorithm has achieved the final result. To apply these algorithms in different fields and test how the methodology works, we need to use eXplainable AI techniques.
Most of the work in the medical field focuses on binary or multiclass classification problems. However, in many real-life situations, such as chest X-rays, radiological signs of different diseases can appear at the same time. This gives rise to what is known as "multilabel classification problems". A disadvantage of these tasks is class imbalance, i.e. different labels do not have the same number of samples. 
The main contribution of this paper is a Deep Learning methodology for imbalanced, multilabel chest X-ray datasets. It establishes a baseline for the currently underutilised PadChest dataset and a new eXplainable AI technique based on heatmaps. This technique also includes probabilities and inter-model matching. The results of our system are promising, especially considering the number of labels used. Furthermore, the heatmaps match the expected areas, i.e. they mark the areas that an expert would use to make the decision.

\end{abstract}

\section{Introduction}
\label{sec:Introduction}

In recent years, the field of medicine has faced two relevant problems that hinder patient care: staff workload and subjectivity in the interpretation of  tests\cite{moustaka2010sources, domingueztesting}. These problems have no easy solution, which is especially dangerous in medicine because procedural errors can lead to serious health complications. Firstly, overwork in medicine, aggravated in recent times by the global COVID-19 pandemic, can lead to errors and delays in diagnosis and treatment. As mentioned above, there is also subjectivity in the interpretation of some medical tests. The expert analysing these tests, for example X-rays, may arrive at an erroneous diagnosis due to, for example, the existence of signs of different diseases to different degrees \cite{shaw1990inter}. This type of imaging test is one of the most common in various diagnoses due to its low cost, speed of acquisition and the fact that it does not require much preparation \cite{ahmed2021discovery}. Chest X-rays are useful for detecting a variety of diseases of the chest related to different organs such as the heart, lungs or bones. The features of X-rays make them suitable for analysis with convolutional neural networks (CNN) \cite{lecun1995convolutional}. The combination of AI algorithms and medical knowledge can improve the performance of medical staff \cite{kontzer} and could also reduce patient waiting times by speeding up the diagnostic process and reducing the workload of doctors.

CNNs have been a breakthrough in computer vision due to their ability to extract features from images. These architectures are composed of different layers. The first has convolutional layers that are inspired by the notion of cells in visual neuroscience. The architectures are based on the visual cortex of animals. The main reason why these architectures have stood out is their great capacity to extract patterns from data, improving the performance of previous systems based on Machine Learning models. This advantage has made them a benchmark in Deep Learning due to their high performance in a wide range of tasks, such as speech recognition, computer vision or text analysis \cite{lecun2015deep}. 

The properties of chest X-rays make them susceptible to be analysed by this type of algorithms. Some of the main advantages of CNNs over traditional techniques are that it is not necessary to manually extract image features or perform segmentation, and that by being able to learn from large volumes of data they can identify patterns that are difficult for the human eye to detect. Although in this article we focus on classification problems, other problems can be solved, such as X-ray segmentation, localisation, regression (such as predicting drug dosage), among others.
CNNs are a potential tool for the analysus of chest radiographs. However, most of the work in this field focuses on binary and multiclass classification problems.
Actual problems are usually more complex than the above; they tend to be multilabel classification problems, i.e. the different labels are not mutually exclusive, whereas in binary and multiclass classification problems there is only one label per radiograph \cite{baltruschat2019comparison}.
To solve multilabel problems, we need to explore new strategies. Adapting algorithms can interpret this kind of problems by transforming them into simpler problems that can be solved by traditional algorithms, i.e., transforming them into binary problems \cite{park2020marsnet}.
In the field of chest X-rays we can find samples without labels, healthy patients and samples with radiological signs of several diseases at the same time.
On the other hand, there are a large number of different radiological signs in chest X-rays, so if we want to build and validate a system that approximates realistic conditions, we have to use a dataset with a large number of mutually non-exclusive labels. This is the case of the PadChest database \cite{bustos2020padchest}, which has 174 different radiological signs, substantially increasing the degree of realism and the complexity of the problem. 
 
Many machine learning algorithms, including CNNs, work best when the classes in the dataset are balanced. However, in real life it is common to find datasets where this condition is not met; they are imbalanced datasets, where one or more classes have substantially more examples than the rest.
As a consequence, with such datasets, machine learning algorithms learn a bias towards the majority class, even though the minority class is often more relevant. Therefore, it is necessary to apply different methods to improve the recognition rate\cite{al2020neuro}.
There are several options to overcome this difficulty: a) modify the dataset, reducing the samples from the majority class or increasing the number of samples from the minority class; b) modify the algorithms to alleviate their bias towards the majority class, e.g. weighted learners \cite{lin2020deep}. The problem of unbalanced databases is exacerbated in multilabel classification problems, where multiple minority classes may appear, making this challenge more difficult to solve.
In medicine, it is widespread because each disease has a different incidence in the population.
Heart disorders top the list of the deadliest diseases, followed by chronic obstructive pulmonary disease, which causes more than 6 million deaths a year.
In contrast, other diseases such as lung cancer are the sixth leading cause of death with less than 2 million deaths, according to the World Health Organization \footnote{https://www.who.int/es/news-room/fact-sheets/detail/the-top-10-causes-of-death}.
As a result, most radiographic datasets are imbalanced; a clear example is PadChest, the dataset used in this article, where the number of samples in each class approximates the incidence published by the World Health Organization.

These algorithms, like many other Deep Learning and Machine Learning methods, are considered "black box" algorithms because end users can only analyse the input and output, but the inference process is opaque, which reduces confidence in these algorithms. To alleviate this problem, explainable AI techniques have been developed, such as saliency maps, which produce heatmaps that highlight the pixels with the greatest influence on the final prediction \cite{mundhenk2019efficient}.
This problem is serious in medicine, where errors can be dangerous for patients \cite{holzinger2022information}.
For this reason, explainable AI techniques are essential, as they allow users to understand how the system has arrived at the final result and use it to help diagnose \cite{singh2021covidscreen}. However, the combination of medical knowledge and AI has many advantages, such as helping to reduce medical errors, and speeding up diagnostic processes, leading to improved patient care, as doctors would have more time to attend patients.

\begin{figure}[!h]
\centering
    \mbox{\includegraphics[width=3.50in, scale=1]{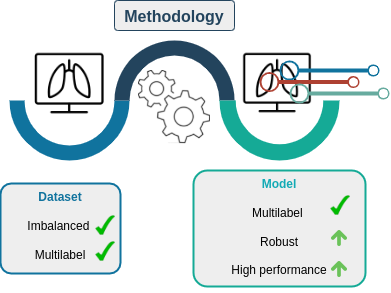}}
  \caption{Visual representation of the problem and the objective of the methodology.}
  \label{fig:diagram_introduction}
\end{figure}

The contribution of this manuscript is a methodology for classifying imbalanced multilabel datasets with many classes, as shown in Figure \ref{fig:diagram_introduction}.
The aim of this methodology is the generation of robust and quality models; in this case, it has been applied to a highly imbalanced multilabel chest X-ray dataset with 174 classes.
The methodology consists of different stages: preprocessing of the images based on segmentation, training of the state-of-the-art models and, finally, a heatmap-based visualisation, which allows the end-user to understand how the model has classified the sample.
We have chosen the PadChest dataset, which consists of more than 160,000 images with 174 different labels. We selected this dataset because of the number of classes, which is larger than in the rest of the state-of-the-art datasets, and the labels are highly imbalanced.
The "normal" label (without radiological signs) represents more than 37\% of the samples, while other classes have fewer than 100 samples, such as breast or pleural cancer.
However, this methodology will allow to establish a suitable benchmark for this dataset against which future work can be compared, as there are currently very few published papers using this dataset and they do not make a detailed analysis of the problem.

We can summarise the main contributions of this research as follows:

\begin{itemize}
    \item A methodology for imbalanced multilabel classification problems.
    \item Experiments using a dataset with a large number of classes (more than 30) and severe imbalance.
    \item Explainability using Grad-CAM for multilabel datasets. 
\end{itemize}

Finally, this manuscript is organised as follows. Section 2 summarises the most relevant work in the literature, with a special focus on chest X-ray classification problems for imbalanced multilabel datasets; section 3 describes the methodology proposed for this type of problem, consisting of training a model and generating a visualisation based on heatmaps; section 4 presents the CNN architectures, the hyperparameters used for training, details of the execution environment, and a link to the repository where the code used in the experimentation can be found; section 5 presents the experimental results, and section 6 presents the main conclusions and possible lines of future work.

\section{Related work}
\label{sec:relatedwork}

Since the first application of AI techniques in medicine in the 1980s, the use of these algorithms has grown exponentially, especially in recent years.
Deep learning algorithms are applied to all kinds of clinical data \cite{piccialli2021survey}: biosignals, which include electrical \cite{baalman2020morphology, chung2019real}, mechanical \cite{li2020fusion, pan2017liftingnet} and thermal signals \cite{su2021deep, abdullah2022multiple}; biomedicine, which studies molecules of biological processes \cite{jost2020titrating, li2020deep, zampieri2019machine}; electronic health records (EHR), focused on optimising diagnosis \cite{welch2020user, xu2020identifying, rough2020predicting}; and clinical imaging, widely used in the diagnosis of many diseases \cite{arefan2020deep,byra2020knee, saha2020predicting}, as is the case with our problem.
The practice of healthcare has evolved from observation-based medicine to evidence-based medicine.
This makes deep learning and big data algorithms especially useful in this field as they can indentify some radiological signs that medical staff cannot detect\cite{goodfellow2016deep}.
Although in this manuscript we focus on classification problems, there are papers where these algorithms are used in regression problems, such as estimating the dose of a drug \cite{kazemifar2020dosimetric}; generating medical reports from clinical tests \cite{liu2019clinically}; or image processing, such as image segmentation \cite{nemoto2020efficacy} and image reconstruction \cite{benz2020validation}.

The COVID-19 pandemic has had a strong impact on research into the application of machine learning and deep learning in medical image analysis.
As expected, many of the classification systems investigated have focused on detecting signs of bilateral COVID-19-associated pneumonia.
In \cite{ahmed2021discovery} they use two different pre-trained architectures to classify chest X-rays, VGG16 and ResNet, and optimise the hyperparameters.
In \cite{pham2021classification} they train three different pre-trained architectures, AlexNet, GoogleNet and SqueezNet, with six datasets independently, testing different percentages of train set samples (50 and 80 \%), achieving an accuracy of 99.85\% with SqueezeNet.
However \cite{ahmad2021deep} develops an ensemble system based on MobileNet and InceptionV3 that achieves 96.49\% accuracy.
Soon, binary classification was extended to multiclass problems, making it possible to discern whether pneumonia is caused by COVID-19 or another virus/bacteria or whether the patient is healthy.
As with binary classification problems, many works, such as \cite{avola2021study}, use state-of-the-art architectures to find the best performing ones, such as AlexNet, GoogleNet, ResNet and ShuffleNet, among others.
MobilNet\_v3 achieves the best result with a precision of 84.92\% on a dataset composed of 6330 samples.
In \cite{zebin2021covid},  in addition to training a pre-trained state-of-the-art architecture, a heatmap-based visualization is generated that shows two images for each sample. The first is the original X-ray, and the second is the class activation map, i.e. the most important area for the CNN.
However, the images do not overlap, making interpretation difficult. Other works, such as \cite{teixeira2021impact}, apply segmentation techniques to remove all irrelevant areas of the system, which should improve performance and visualization.
Their dataset consists of three different classes: COVID-19, normal and lung opacity.

As we have discussed in Section \ref{sec:Introduction}, the explainability of deep learning models is a fundamental factor to be taken into account in their application.
These models are black-box algorithms and need explainable AI techniques to make them more trustworthy \cite{arrieta2020explainable}.
There are two main ways to produce the final visualization, 1) generate a heatmap per label, or 2) generate a single visualization for all classes.
The first one is more commonly used, \cite{teixeira2021impact, avola2021study, wang2017chestx} however, this technique has one main limitation: it is not feasible for a large number of labels, and it makes a global view difficult.
The second one (e.g. \cite{teixeira2020dualanet}), shows the different signs as areas with a higher colour intensity, but only one colour scale was used, which makes it difficult to identify which pathological sign indicates which area of interest. We propose a new technique where each visualisation shows a radiological sign including the probability and agreement between models.

Although most medical datasets have two classes (samples of a particular pathology and healthy samples), in chest X-rays it is common to find signs of more than one pathology.
For this reason, in the last five years different authors have published multilabel radiological datasets. These datasets are closer to real situations than binary ones, with the additional challenge of imbalance of different classes. The size of each class in a realistic dataset should depend on the incidence of  pathology in society, i.e. some classes are more represented than others. These characteristics of these datasets are interesting and need to be analysed in detail in order to address the problem adequately.

\subsection{Multilabel classification problems}
\label{subsec:multilabel}
As we have seen, much of the work generated in recent years has focused on binary and multiclass classification problems. In these problems, the labels are mutually exclusive, while multilabel classification problems have multiple classes that are not mutually exclusive, which increases the difficulty of the problem. 
There are two ways to solve these problems: a) transform the multilabel problem into simple binary problems, or b) adapt the algorithms to solve the multilabel problem directly, i.e. attack the problem globally \cite{allaouzi2019novel}. 

Binary and multiclass classification systems are very restrictive, as they only serve to detect one type of radiological finding.
However, patients can often present signs of multiple diseases at the same time.
There are very few multilabel datasets that take into account a large number of signs, as they require a large number of samples and most of them have only a few labels.
Three datasets are worth highlighting for their quality and relevance to the state of the art (see Table \ref{tab:datasets}).
The first two have been used extensively in image classification problems, but the third has been used mainly in medical report generation \cite{monshi2021labeling, liu2019clinically, smit2020chexbert, boag2020baselines, jain2021visualchexbert}.
However, this third dataset has two advantages that make it very suitable also for classification problems: 1) the number of labels is larger; 2) it has the largest number of different patients, which implies a smaller number of similar samples from the same patient.
Given the lack of application of algorithms for this task to increase the potential and interest of this dataset mentioned above, it was selected as a case study for this article.

\begin{table}[h]
\caption{Summary table of multilabel datasets in the field of chest radiography.}
\centering
\begin{tabular}{r|lllll}\toprule
              & \# samples & \# patients & \# labels & views           & reference \\ \midrule \midrule
ChestX-ray 14 & 112120     & 32717       & 14     & frontal         &    \cite{wang2019chestx}       \\
CheXpert      & 224316     & 65240       & 14     & frontal/lateral &    \cite{irvin2019chexpert}       \\
PadChest      & 160000     & 67000       & 174    & frontal/lateral &    \cite{bustos2020padchest}      \\ \bottomrule
\end{tabular}
\label{tab:datasets}
\end{table}

ChestX-ray 14 dataset is one of the most widely used datasets in the field of chest X-ray classification since its publication in 2019 \cite{wang2019chestx}.
For example, \cite{wang2020detecting} uses DenseNet-121 optimising its hyperparameters, obtaining an average AUC of 0.82. The AUC achieved for the pneumonia class was 0.662 (the lowest), while for the hernia class it was 0.923 (the highest).
Other researchers use different architectures such as InceptionResNet\_v2 and ResNet152\_v2 to achieve an AUC for pneumonia of 0.73 \cite{albahli2021ai}.
Much of the work on this dataset retrains state-of-the-art architectures, but there are other strategies for improving classification performance; for example, \cite{almezhghwi2021convolutional} switches the classifier from AlexNet and VGG16 to SVM with the intention of improving the results of previous manuscripts, achieving an AUC for pneumonia of 0.98 with both architectures.
Having different types of radiographs of the patient can also improve the classification results, for example a frontal and a lateral X-ray.
Finally, the main disadvantage of ChestX-ray 14 dataset is that it only contains radiographs with a frontal view, while CheXpert and PadChest datasets also contain X-rays with a lateral view.

The second dataset, \textbf{CheXpert} has 14 different labels and reports on all images.
In terms of published classification work, we find a situation similar to ChestX-ray 14, with many works retraining state-of-the-art architectures, such as \cite{seyyed2020chexclusion}, where they adjust the hyperparameters of DenseNet-121 to optimise its performance.
Other authors look for different strategies, such as \cite{cohen2020limits}, where they make two modifications to DenseNet to improve its performance.
First, they modify the loss function by assigning weights to the different labels, alleviating the imbalance problem. Second, they modify the threshold for discerning between the presence or not of each label, i.e. the probability at which the class is considered present.
However, the CheXpert dataset has the same limitation as ChestX-ray 14: they contain only 14 possible diseases, which represents only a small subset of all possible diseases that may be present in the chest.

Finally, \textbf{PadChest} is the most interesting dataset of the three in our opinion because it has many more labels than the others. It is a massive multilabel classification problem, much closer to reality than the other datasets.
The number of patients used is also larger than in the others, leading to more variability in the dataset, and the imbalance of the classes is larger too.
One of the papers using this dataset, \cite{hashir2020quantifying}, combines the PA and lateral views to predict labels in four different ways: a) the lateral view is stacked in the second channel of PA X-ray; b) both views are processed by two CNNs and the combination of them is processed by a fully connected layer; c) the model input is processed through two separate CNNs, the output is concatenated and passed by two dense layers with an average pooling layer between them; d) a modification of c) where two dense layers are added.
A major limitation of that paper is that it shows overall results without performing a detailed analysis per label, which prevents comparison with other works in the area.
On the other hand, in \cite{pooch2020can} CheXNet is retrained, which is a state-of-the-art architecture previously trained with a multilabel chest X-ray dataset.
In that paper, different models are trained with four datasets, and each model is tested with each dataset separately.
The main limitation of that manuscript is the reorganisation of the labels of PadChest dataset: the label "Lesion" is generated to unify the samples of the atelectasis classes, using only 8 classes out of the 174 available.
Given the limitations we have found in all classification works using the PadChest dataset and that some most of them are not replicable, we propose to create a benchmark that future works can use to compare results, with a methodology adapted for the two main problems: the high number of different labels, and the imbalance between them. We propose two ways of splitting the dataset based on the term tree provided by its authors, which allows us to group radiological signs into higher classes. The first uses the specific labels for a finer-grained classification, and in the second division we will work with the higher labels, which indicate more general radiological signs.

\subsection{Class imbalance in Deep Learning}

As explained above, most machine learning algorithms work best when the number of samples for each class is similar.
When there is a significant difference between the classes, the system will boost the majority class while the minority class(es) will have less relevance, even though the minority class is often the most relevant. 
There are several classification tasks with this problem, such as \cite{cohen2020covid}, where the majority class is COVID-19 over the rest of the pneumonia classes.
As expected, because the incidence of COVID-19 has been extremely high, the dataset contains more than 400 samples of COVID-19 followed by the class \textit{Pneumocystis spp} with fewer than 30 samples. This phenomenon appears in many classification tasks, especially those with more than two classes, both multiclass and multilabel.
For example, in \cite{wang2019chestx} there are 15 classes and the class "No findings"/"Normal" exceeds 50,000 samples, while the other labels have less than 20,000 samples, of which only three exceed 10,000 samples.

As mentioned in the Introduction section, there are different strategies to alleviate the class imbalance problem.
\textit{Modify the dataset}, for example with oversampling techniques, which increase the number of samples from minority classes by applying data augmentation and histogram equalisation techniques \cite{wang2019enhanced}.
\cite{charte2015mlsmote} develops a new algorithm, Multilabel Synthetic Instance Generation, for multilabel problems.
For each sample, a nearest-neighbour search is performed, the features are extrapolated and the label is generated from them.
Another option for generating synthetic samples is to use generative adversarial networks (GANs), i.e. to use deep learning models to produce new samples from the original dataset. \cite{salehinejad2018synthesizing} uses this method to generate new chest X-rays to balance the different classes.
Another strategy for balancing the classes in the dataset is to reduce the samples of the majority classes. This technique is called undersampling.
Tipically, random samples are removed from the majority classes, as in \cite{qu2020assessing}, where the maximum number of samples in each class is set to balance it.
Undersampling is not as widespread as oversampling because Deep Learning systems need a large number of samples, so undersampling may not work.

Another strategy to alleviate class imbalance is to \textit{modify the way the model learns} by increasing the weight of minority classes in learning, thus preventing the model from giving more importance to majority classes.
One option is to apply class weights in the loss function that increase the relevance of the minority classes.
One example is \cite{rajpurkar2017chexnet}, which uses the chest X-ray14 dataset to classify the presence or absence of pneumonia.
Another example is \cite{monowar2020lung}, where the weighted binary cross-entropy loss function is applied. \cite{ge2018chest} developed a novel error function, Multilabel Softmax Loss, this method considers the relationship of multiple labels explicitely, the author compute the derivative of the error with respect to each class using the chain rule. In addition they applied it to a system composed of two CNNs combined by a bilinear pooling layer. \cite{teixeira2020dualanet} proposes a dual lesion attention network composed of two models, DenseNet-169 and ResNet-152, as feature extractors, after an attention module and average max pooling. The outputs are combined to generate three classifiers. Finally, all classifiers are merged to obtain the final prediction. In addition, they used a variant of the weighted binary cross-entropy loss. To tackle the class imbalance, we propose using weighted cross-entropy with logits using class weights.

\subsection{The challenge of imbalance in multilabel classification problems}

As we have explained, many real classification problems have two properties that make them difficult to solve: multilabeling and imbalance.
Each of these two properties alone makes classification difficult, so together they can be very challenging.
In medicine, multilabel and imbalance problems are common because medical staff can find different radiological signs on a chest X-ray, and different diseases do not have the same incidence in the population.
All the datasets mentioned in Section \ref{subsec:multilabel} have both features; however, ChestX-ray 14 and CheXpert have a low number of classes, 14, compared to PadChest \cite{bustos2020padchest}, which is composed of 174 different labels with a large imbalance: the label "Normal" has more than 35000 samples, while other labels, such as round atelectasis, pleural mass or nephrostomy tube, have less than ten samples. 

Most of the published work using these datasets modifies the architecture so that it can directly solve multilabel problems, but does not consider or apply any specific technique to solve the imbalance problem.
However, other works explore different ways to overcome these difficulties and achieve better results.
Such as \cite{huang2019diagnose}, which proposes a multi-attention convolutional neural network to reduce the performance difference between classes and, more interestingly, to extract discriminative features to classify similar classes, which is very common in this kind of dataset.
\cite{wang2019kgznet} generates three images: the first one is the original chest X-ray, the second one is a segmentation-based cropping, where areas not interesting for the model are removed, and the last one is a cropping of the area where previous models have found pathological signs. The information extracted from the three images is fused and finally processed to obtain the final result.
Another interesting strategy is the modification of the loss function to focus on the most interesting samples; for example, \cite{qin2018weighted} proposes a loss function called "weight focal loss", which forces the model to pay more attention to the most difficult samples. This makes the model pay more attention to minority classes, avoiding false negatives. 

These methods can help in class imbalance problems, but in extreme cases of multilabel and imbalance, such as the PadChest dataset, they may nor be sufficient.
For this reason, we propose a new methodology to alleviate both challenges by generating an ensemble system from different pre-trained state-of-the-art models with a specific loss function for imbalanced datasets, the weighted loss function.
The combination of models dampens the effect individual model errors, while the loss function forces the system to focus on minority classes.

\section{Methodology}
\label{sec:Methodology}

\begin{figure}[H]
\centering
    \mbox{\includegraphics[width=5.0in, scale=1]{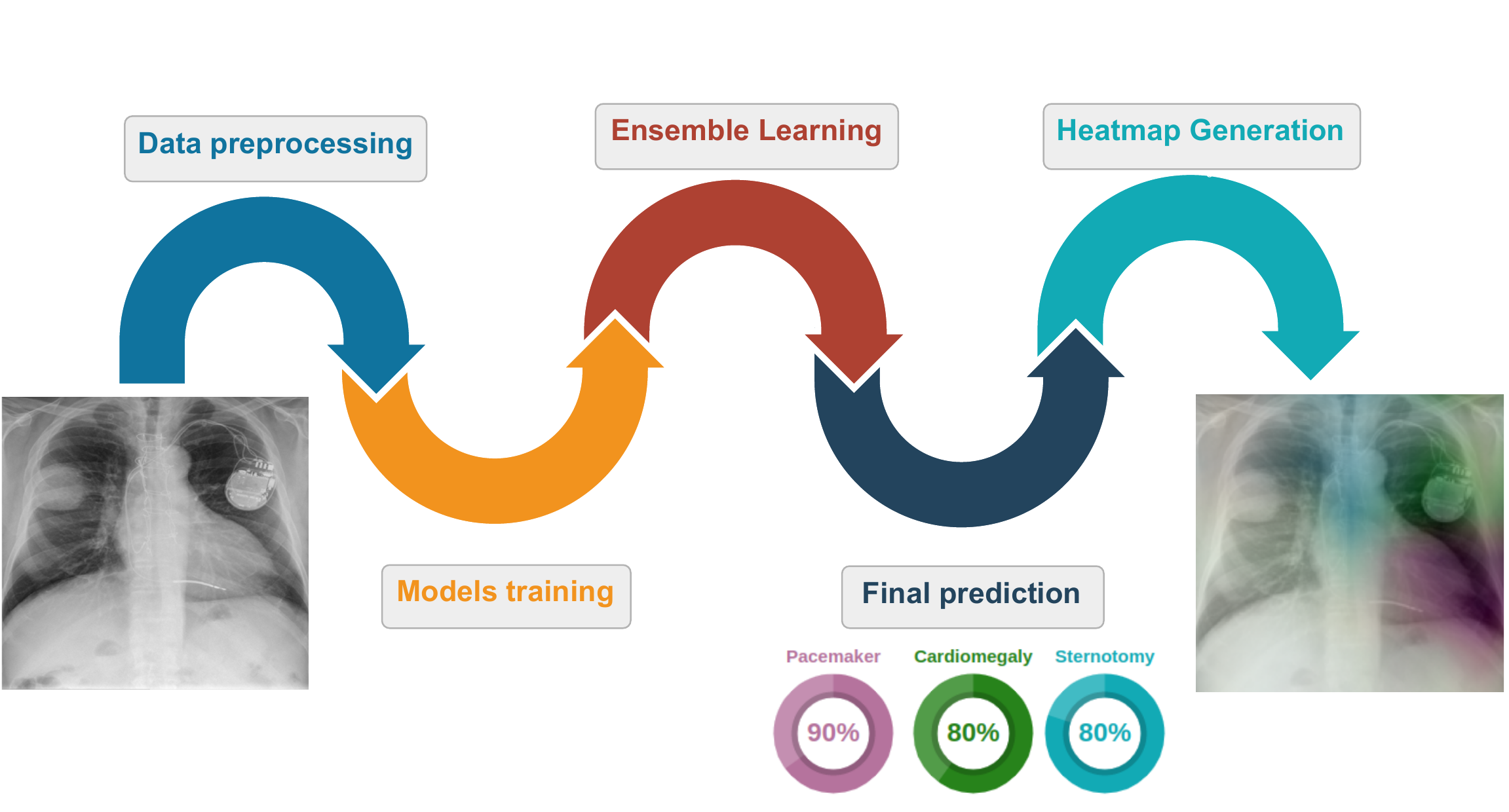}}
  \caption{Visual representation of the proposed ensemble system. We train each architecture with preprocessed images, and all outputs are combined to generate the ensemble. Finally, the system produces the final prediction and heatmap visualisation.}
  \label{fig:diagram}
\end{figure}

We can summarise the proposed methodology in Figure \ref{fig:diagram}, which has four sections.
The first is the data pre-processing step, where we prepare the images for the model and apply data augmentation to alleviate class imbalance.
In the second stage we build the model, training different state-of-the-art architectures.
We then combine the results of each model to obtain the final probabilities.
Finally, we developed a multilabel heatmap technique to areas of the image that are relevant in the classification.
In this technique, the original X-ray is combined with one or more regions labelled with different colours to facilitate the application of these techniques in health centres or hospitals.

\subsection{Label selection}
\label{subsec:labelSelection}

As explained above, multilabel datasets are often imbalanced as they have classes with a low number of samples.
For this reason, we must establish a criterion for choosing the labels to include in our classification system, especially in datasets where the number of classes is extremely high, as in our case.
First, we set the minimum number of samples a label must have to be included in the classification problem, and we set the threshold at 200 X-rays.
For a dataset of 90000 samples this is 0.22\% of the total.
The model cannot work correctly for under-represented labels as it is not a few-shot system.
If a sample has only deleted minority labels we will remove it.

In this paper we consider two different experiments.
First, we use the classes proposed by the authors of the dataset that correspond to the specific labels; this classification system has a smaller number of samples and labels due to the cleaning of under-represented labels explained in the previous paragraph, but is a more fine-grained classification system.
In the second case, we use more general labels. We create these classes grouping the specific classes according to their characteristics. The number of samples and classes is larger at the cost of being less precise systems, but it allows us to cover a larger number of different classes. 

\subsection{Preprocessing}
\label{subsec.preprocessing}

The raw images were preprocessed in order to train the model efficiently.
First, we reduced the number of channels to one as the original samples are RGB images.
Next, we normalised their size to 512x512 pixels.
The pixel values were then normalised between 0 and 1.

The chest X-rays show a larger area than the area of interest (ROI).
Areas such as the arms or neck, among others, are irrelevant to the problem we want to solve, so a cropping based on segmentation masks was performed, forcing the system to focus on the relevant areas.
This trimming is performed in three different steps: first, we generate the lung masks using a segmentation model based on the U-Net architecture \cite{islam2018towards}.
We also added the area underneath the lungs to the masks as it may contain radiological signs of interest.
On many occasions, segmentation models are not perfect; they generate more than two masks, leave gaps inside the masks, etc.
Therefore, thirdly, we decided to use a mask post-processing system \cite{reza2020transresunet}.
This system fills possible gaps in the masks by applying the flood fill algorithm.
Then, if more than two masks have been generated (one per lung), those with an area less than a predetermined value are removed.
In addition, in case the lung masks are stuck together, they are separated. Finally, the image is cropped using the mask coordinates and the lower boundary of the sample.

\begin{center}
\begin{figure}[!h]
    \mbox{\includegraphics[width=\textwidth, scale=1]{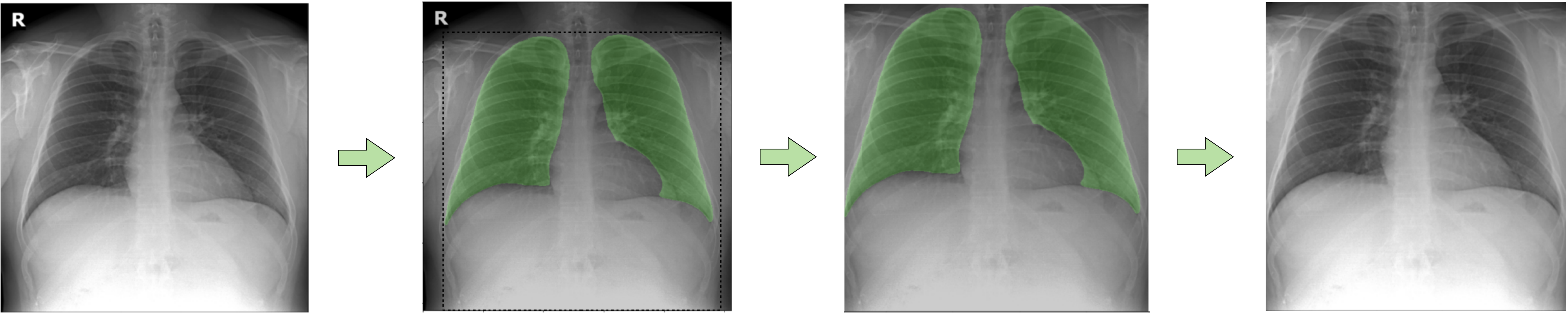}}
  \caption{A segmentation-based cropped sample. The first image corresponds to the original X-ray. The second shows the lung segmentation mask. The third one show the cropped image with lung mask, and finally the last image shows the input of our system, the preprocessing result.}
  \label{fig:crop}
\end{figure}
\end{center}

\subsection{Image classification with CNNs}
\label{subsec.cnn}
Five state-of-the-art architectures pre-trained with ImageNet were selected for their relevance:

\begin{itemize}
    \item EfficientNetB0 \cite{tan2019efficientnet}: This architecture uses different scaling coefficients to scale width, depth and resolution.
    In the EfficientNet family, this architecture is the smallest.
    It is based on the idea that if the images are larger, the network needs more layers to extract the relevant information.
    \\
    \item DenseNet-201 \cite{huang2017densely}: Instead of adding more layers to the architecture, the number of connections between units is increased by connecting each unit to the last, unlike ResNet50, which only connects one unit to the next output.
    This architecture has several advantages: it alleviates the vanishing gradient problem, enforces feature propagation and feature reuse, and reduces the number of parameters.
    \\
    \item InceptionV3 \cite{szegedy2016rethinking}: This architecture is different from the previous ones. It factorises convolutions into smaller convolutions (which can be asymmetric) to reduce cost. In addition, this architecture has an auxiliary classifier between layers that acts as a regulariser. 
    \\
    \item InceptionResNetV2 \cite{szegedy2017inception}: This architecture combines ResNet and InceptionV3. It consists of several Inception units with shortcut connections between them; this enhances the capability of the architecture. 
    \\
    \item Xception \cite{chollet2017xception}: It consists of depth-wise separable convolutions involving two steps: depth-wise convolution, which differs from the standard convolution in that it only acts on one channel; and point-wise convolution, where a 1x1 convolution is applied to all channels. This architecture also includes shortcut connections, such as ResNet50.
\end{itemize}

We applied Transfer Learning on the above five arquitectures and retrained them with PadChest dataset, replacing the classifier in all cases with two dense layers.
We froze the first 10\% of the convolutional layers, as they detect basic patterns and do not need to be retrained.
The remaining convolutional layers are retrained to learn patterns specific to our problem.
The main relevant training parameters are summarised in Table \ref{hyperparameters}.
In addition, a checkpoint is used to save the best model using the validation loss.
Finally, an early stopping algorithm was used to finish training when the validation loss did not improve over the last 25 epochs by more than a threshold of 0.001.

\begin{table}[h]
\caption{Summary of parameters used in training: optimization, data augmentation and training methodology.}
\label{hyperparameters}
\centering
\begin{tabular}{rll}\toprule
\textbf{Optimization}               &                                                                               &   \\ \midrule
Optimizer                        & Adam                                                                          &  \\
Learning rate                    & 1e-4                                                                          &  \\
Loss                             & \begin{tabular}[c]{@{}l@{}}weighted cross \\ entropy with logits\end{tabular} &  \\
 \\ 
\textbf{Feed-forward classifier} &                                                                               &   \\ \midrule
\# Neurons                       & 512                                                                           &  \\
Activation                       & ReLu                                                                          &  \\
Dropout                          & 0.2                                                                           &   \\ \\
\textbf{Data Augmentation}       &                                                                               &   \\ \midrule
Shear range                      & 0.1                                                                           &  \\
Zoom range                       & 0.1                                                                           &  \\
Rotation range                   & 45                                                                            &  \\
Width shift range                & 0.1                                                                           &  \\
Height shift range               & 0.1                                                                           &  \\
Horizontal flip                  & True                                                                          &  \\
Fill mode                        & nearest                                                                       &  \\
Brightness range                 & 0.7-1.1                                                                       &  \\
Channel shift range              & 0.05                                                                          &   \\ \\ 
\textbf{Training methodology}    &                                                                               &   \\ \midrule
Maximum epochs                   & 350                                                                           &  \\
Early stopping patience          & 25                                                                            &  \\
Early stopping threshold         & 0.001                                                                         &  \\
Batch size                       & 32                                                                            &  \\
Image size                       & 224x224                                                                       & \\ \bottomrule
\end{tabular}
\end{table}

\subsection{Ensemble technique}

Ensemble learning is an effective way to improve the performance and robustness of deep learning algorithms.
We combined the results of all trained models, obtaining a system composed of five different architectures with the same test set.
We distinguish two approaches \cite{nguyen2020aggregation}: "Combine then predict" (CTP) and "Predict then combine" (PTC).
In the CTP method, the label probabilities predicted by the individual models are first calculated, and then the average probability at each label is used to obtain the ensemble label prediction.
The other method, PTC, combines the binary predictions to obtain the ensemble.
We consider two versions of PTC: label-wise voting (PTC-lw), which calculates the number of positive and negative individual predictions for each label, adopting the majority.
Thus, PTC-lw calculates the prediction of each label independently of the others.
On the other hand, PTC-mode calculates the set of labels predicted by each individual model, and predicts the most frequent set.

\subsection{Heatmap generation}

As explained in Section \ref{sec:relatedwork}, the output based on the ensemble is limited for medical staff because they cannot understand how the system obtains the results.
For this reason, we developed a visualisation technique using heatmaps.
This technique uses a colour scale to highlight the pixels in the original image that are most relevant to the model.
We use a grad-CAM algorithm to visualise the activation maps of the classes. First, we change the activation function of the classifier to linear and use Tensorflow to generate the class activation maps, which calculate the gradient of the image with respect to the activation of the convolutional layer.
We then compute the gradient of the output neuron with respect to that obtained from the convolutional layer to indicate the areas where the system is focused. Finally, to generate the ensemble heatmaps, we obtain the activation maps of all neurons and compute the average value of each pixel for each class.
Each heatmap is overlaid on the original X-ray using a 10\% transparency. The title includes the probability for this class and the inter-model agreement which shows the confidence of the ensemble in that prediction.

\section{Experimental setup}
\subsection{Dataset}

In this article, we have used the PadChest dataset \cite{bustos2020padchest}, an imbalanced, multilabel dataset.
It was published in January 2019 by the University of Valencia together with BIMCV.
The samples were collected at Hospital de San Juan (Spain) between 2009 and 2017.
This dataset is composed of 160,868 clinical images from 67,625 patients, divided into 174 different labels, and corresponds to different signs of thoracic disease. This dataset contains chest X-rays with different projections: posteroanterior (PA), anteroposterior (AP) and lateral views; however, only PA X-rays were used for experimentation, corresponding to 91,728 clinical images from the original dataset.
The authors of the dataset provided a term tree \footnote{https://github.com/auriml/Rx-thorax-automatic-captioning} in which all labels are grouped into more general classes, as can be seen in Figure \ref{fig.treeTerm}.
In this example, the general class is fracture. The specific labels are clavicle fracture, humeral fracture, vertebral fracture, and rib and callus rib fractures.
Therefore, we designed two experiments, the first using specific classes for classification and the second using more general classes that include several specific classes.
We then set the minimum number of samples that each class must have to be included in the classification system.
The more general classification system has a larger number of classes that are more heterogeneous, while the more specific classification system has a smaller number of classes, but is  more precise than the previous one.
Table \ref{tab:dataset_details} shows the details of the two classification systems, the number of samples, the classes and the size of the training, validation and test sets. In the train/test/validation split we stratify the samples according to classes and patient id, which avoids biases and problems between subsets. In addition, to facilitate the replicability and transparency of this article we will make the split available on the github in section \ref{subsec:execution_environment}.

\tikzstyle{every node}=[draw=black,thick,anchor=west]
\tikzstyle{selected}=[draw=blue]
\tikzstyle{optional}=[dashed,fill=gray!50]
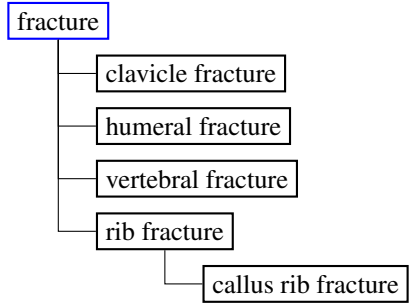
\begin{figure}[h]
\centering
\begin{tikzpicture}[%
  grow via three points={one child at (0.5,-0.7) and
  two children at (0.5,-0.7) and (0.5,-1.4)},
  edge from parent path={(\tikzparentnode.south) |- (\tikzchildnode.west)}]
  \node [selected]{fracture}
    child { node {clavicle fracture}}		
    child { node {humeral fracture}}
    child { node {vertebral fracture}}
    child { node {rib fracture}
      child { node {callus rib fracture}}
    };
\end{tikzpicture}
\caption{Example of a section of the term tree of the dataset. The general class is boxed in blue, and the specific classes are marked in black.}
\label{fig.treeTerm}
\end{figure}

\begin{table}[H]
\centering
\caption{Summary table of the two types of experiments performed (general classes, and specific classes). The total number of classes and the total number of samples in each of the splits are shown.}
\label{tab:dataset_details}
\begin{tabular}{r|lllll}
 \toprule
                 & \# classes & \# samples & train size & val. size & test size \\ \midrule\midrule
General classes  & 54         & 90687      & 63475      & 9069      & 18143     \\ 
Specific classes & 35         & 85367      & 59753      & 8532      & 17082     \\ \bottomrule
\end{tabular}
\end{table}

\paragraph{Label distribution: Specific classes}

\begin{figure}[H]
\centering
    \begin{subfigure}{0.5\linewidth}
\includegraphics[width=\linewidth]{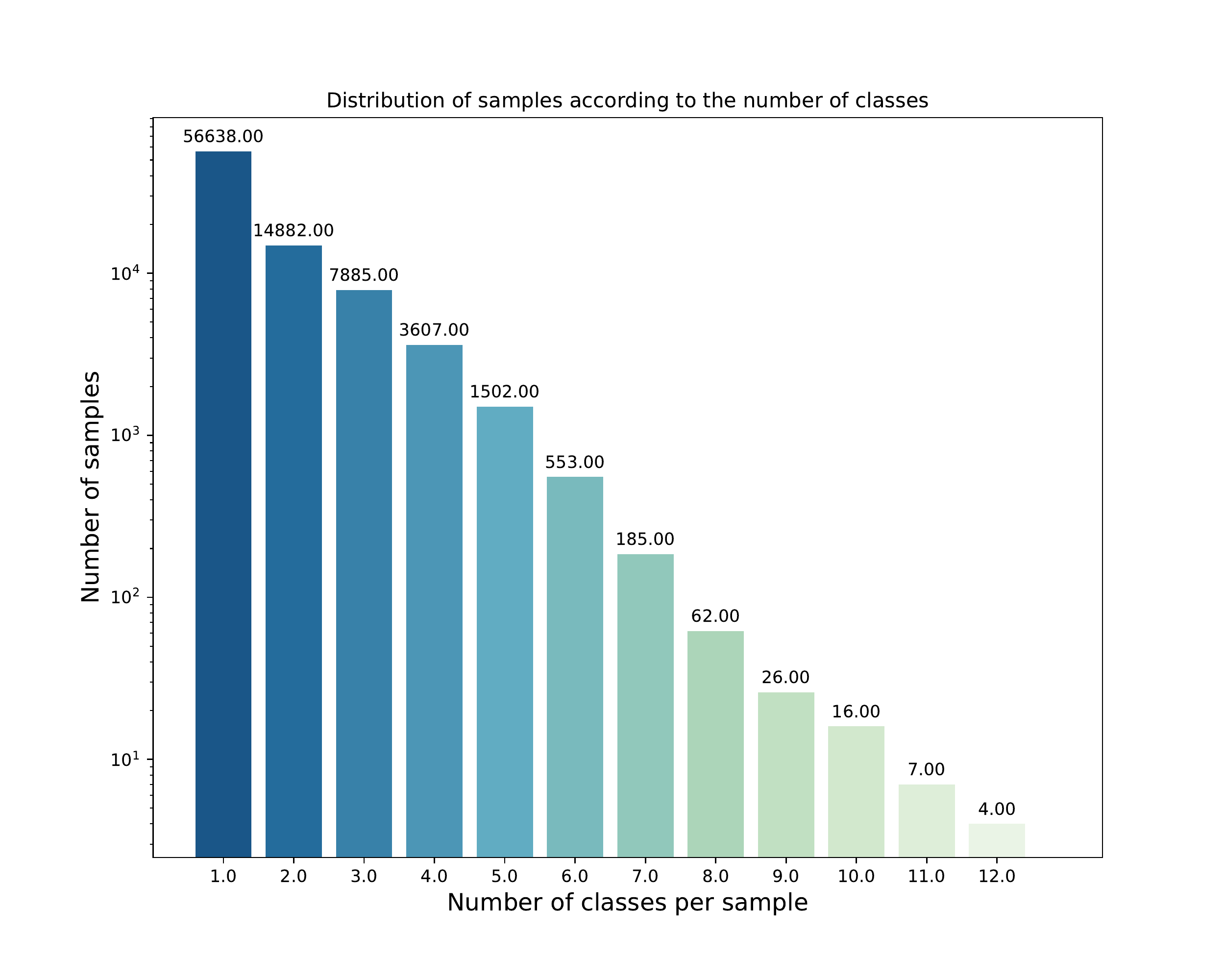} 
    \caption{Distribution of the samples according to the number of labels present in each sample (specific classification experiment).}
\label{fig:specificClassesDistribution}
    \end{subfigure}\hfill
    \begin{subfigure}{0.5\linewidth}
\includegraphics[width=\linewidth]{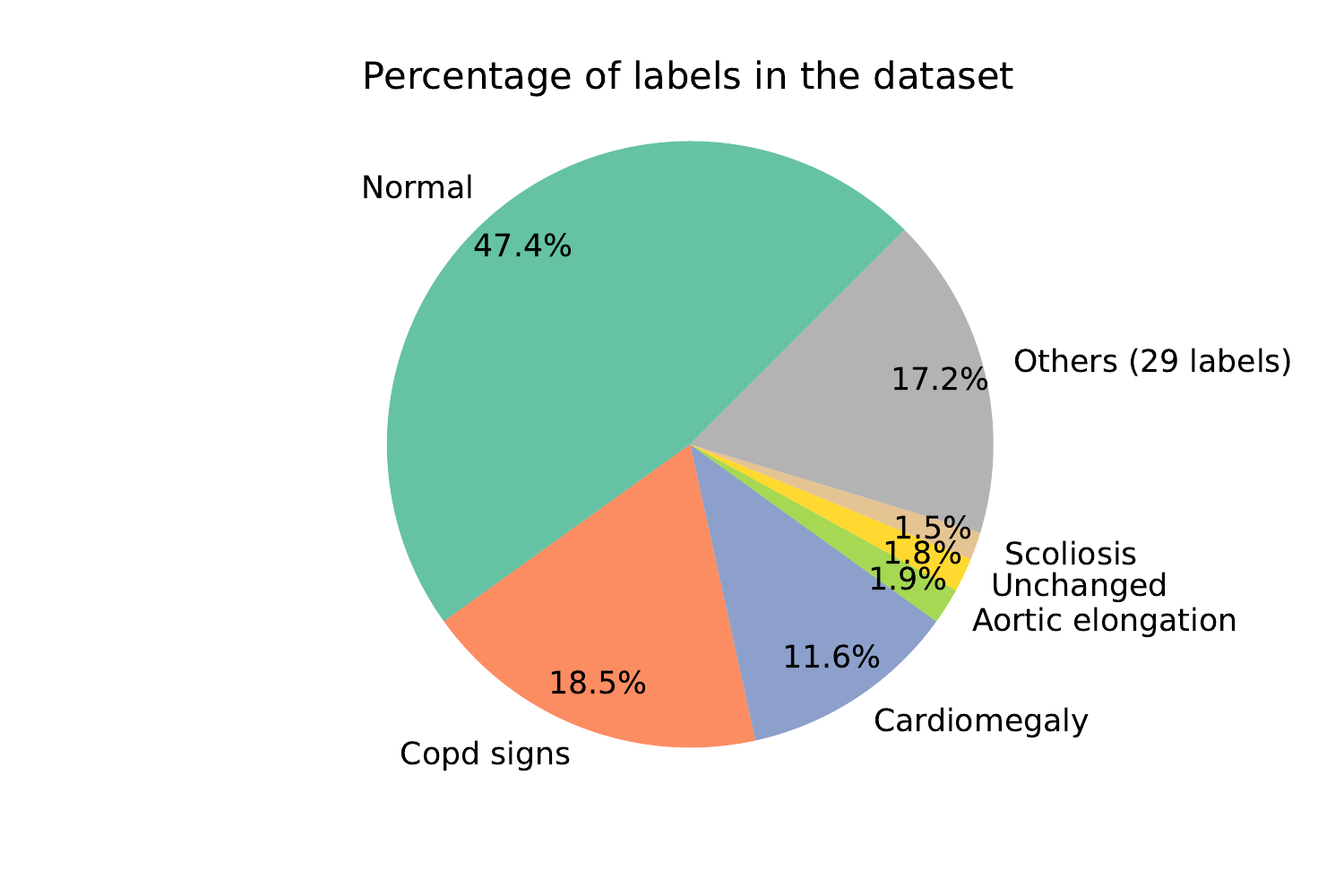}
    \caption{Percentages of samples in each class of the dataset.}
\label{fig:pieplot_v1}
    \end{subfigure}
\caption{Distribution of specific classes. Visual representation of the characteristics of the multilabel and imbalanced training set.}
    \label{fig:featureSpecificClasses}
    \end{figure}

This experiment, as explained in Section \ref{subsec:labelSelection}, label selection, has a smaller number of samples and classes than the second case, but the radiological signs are more accurate. 
In this experiment we used a total of 85367 samples and 35 different classes.
We can observe in Figure \ref{fig:specificClassesDistribution}, how more than half of the samples present a single class; however, we can observe that there are samples with a high number of classes, four of them presenting 12 different labels at the same time.
This distribution of the samples is in line with expectations; the number of samples decreases as the number of labels per sample increases.
In Figure \ref{fig:pieplot_v1} we can see how the classes in this experiment are extremely imbalanced. Although there are 35 classes, the six majority classes account for 82.7\% of the dataset.
Only the normal class, which is the majority class, accounts for 47.4\% of the total samples, while the supra aortic elongation class, which is the least represented class, accounts for only 0.28\% of the total.

\paragraph{Label distribution: General classes}

\begin{figure}[!h]
\centering
    \begin{subfigure}{0.5\linewidth}
\includegraphics[width=\linewidth]{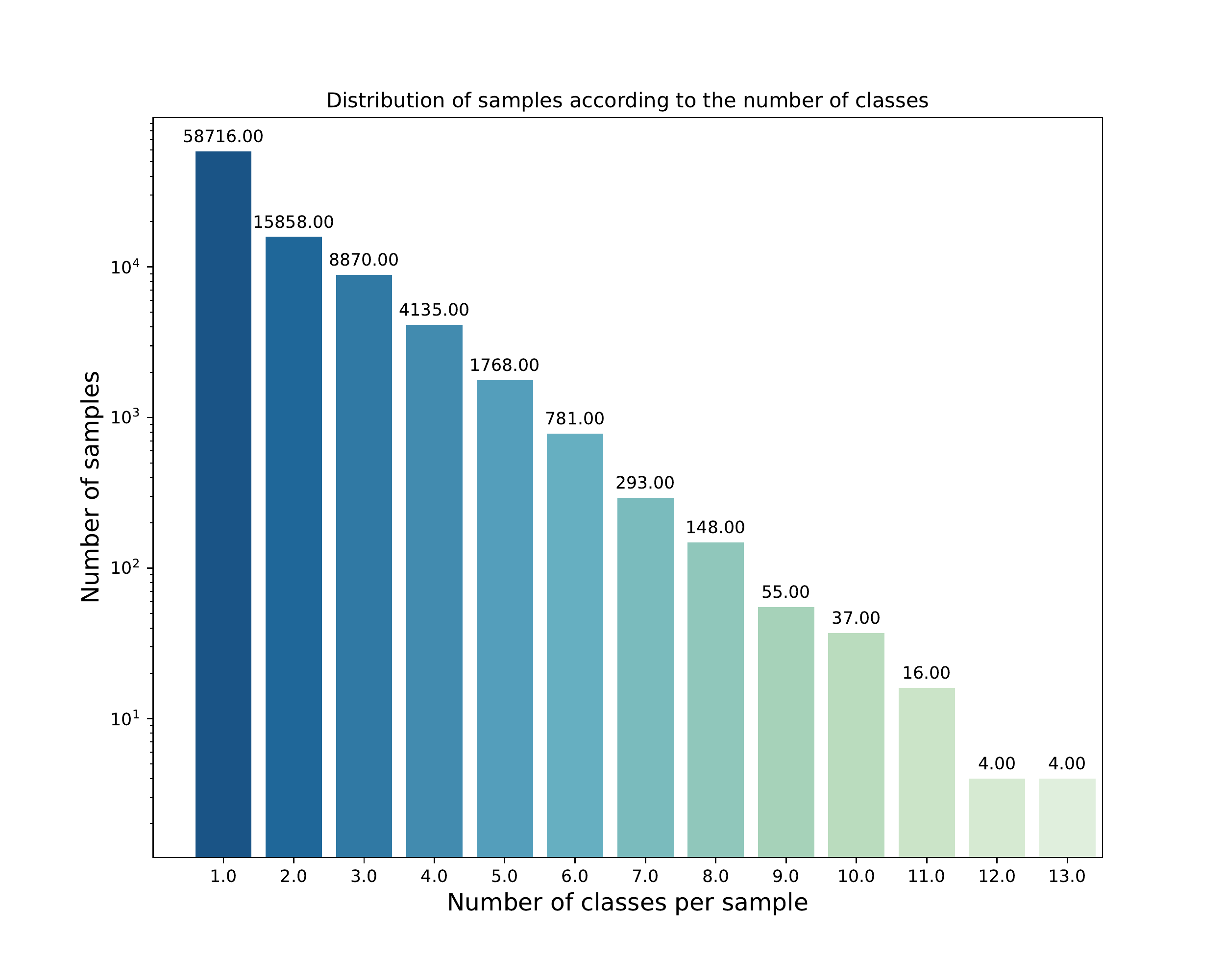} 
    \caption{Distribution of the samples according to the number of labels present in each sample (general classification experiment).}
\label{fig:generalClassesDistribution}
    \end{subfigure}\hfill
    \begin{subfigure}{0.5\linewidth}
\includegraphics[width=\linewidth]{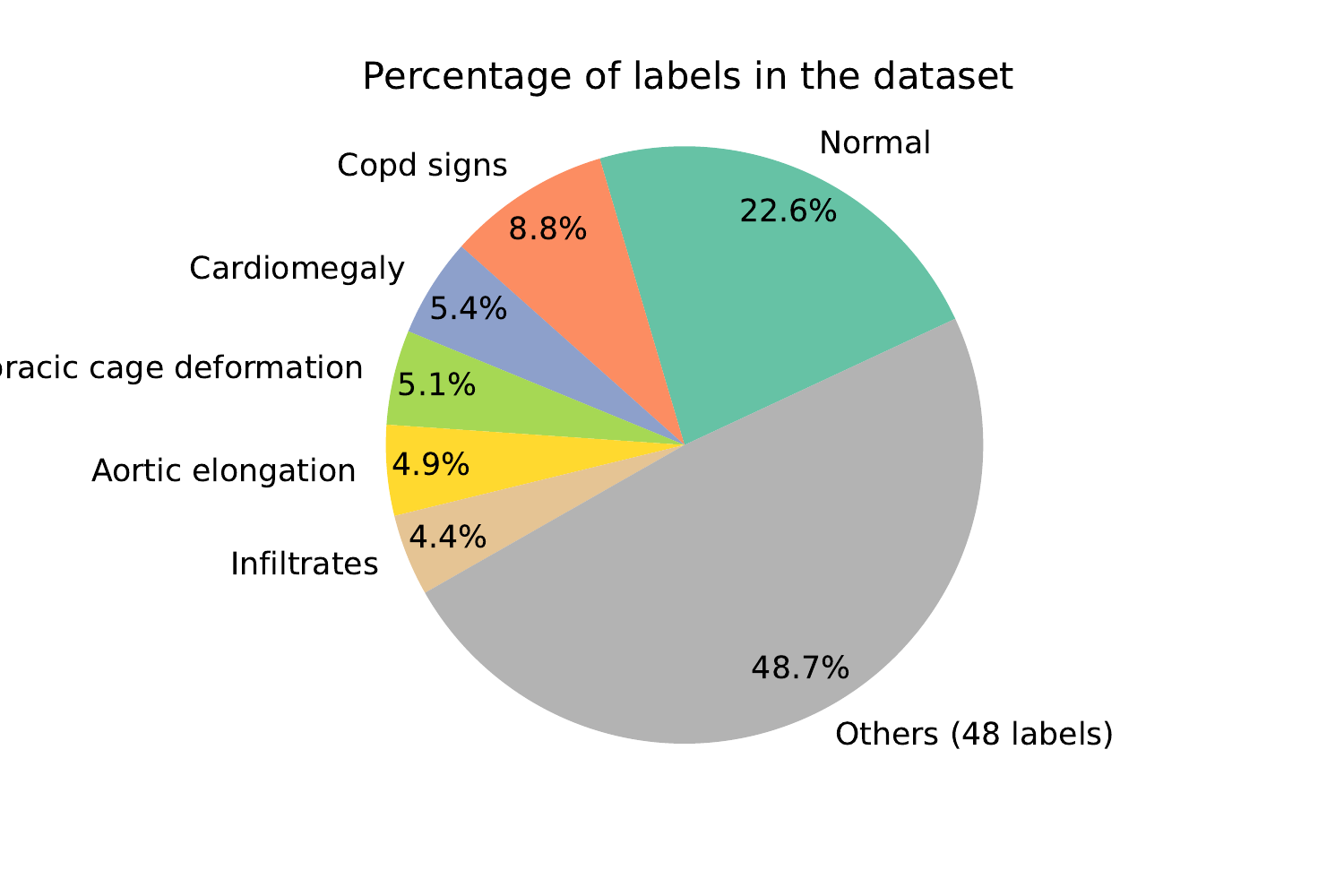}
    \caption{Percentages of samples in each class of the dataset (general classification experiment).}
\label{fig:pieplot_v2}
    \end{subfigure}
\caption{Distribution of the general classes. Visual representation of the characteristics of the multilabel and imbalanced training set.}
    \label{fig:featureGeneralClasses}
    \end{figure}

In this experiment, different classes were unified according to the tree of terms proposed by the authors. Therefore, the number of classes and samples is higher than in the first experiment. However, the radiological signs used in the classification are less precise, so in the end 54 classes and 90,687 samples were used.
In Figure \ref{fig:generalClassesDistribution} we can see how the number of classes per sample is distributed in a very similar way to the previous case.
However, we can see that there are samples with 13 different labels, one more than in the previous case.
If we look at figure \ref{fig:pieplot_v2}, we can see that the six majority classes represent 51.3\% of the total, while the other 48 classes do not reach 50\%.
The majority class, as in the previous case, is the normal class. This class accounts for 22.6\% of the total while the minority class, vascular redistribution, accounts for only 0.13\% of the dataset.
This shows that even if we group the radiological signs into higher classes, the dataset is very imbalanced.

\subsection{Execution environment and Github repository}
\label{subsec:execution_environment}
All experiments have been run on a 24GB Nvidia GeForce RTX 3090. The main packages used in these experiments are the following: Tensorflow \cite{abadi2016tensorflow}, Scikit-Learn \cite{scikit-learn} and openCV \cite{opencv_library}.
The code developed in our work is publicly available at https://github.com/helenalizlopez/PadChest.

\section{Experimental results}
\label{sec:result}
This section describes the results obtained with the proposed methodology and evaluates its performance on a multilabel and imbalanced problem, the PadChest dataset.
We considered two strategies for the classes: directly using the labels proposed by the dataset creators, or grouping them into more generic classes that encompass similar radiological signs.
First, we checked whether preprocessing improves the ensemble performance.
Next, we checked the performance of both the individual models and the ensemble, and analyse the quality of the visualisations based on explainable AI techniques.
To measure the performance of the different models, we have used three metrics suitable for multi-label problems: Area Under the Curve (AUC), Hamming Loss and F-measure \cite{charte2019remedial}.

\subsection{Impact of preprocessing techniques}

\begin{table}[H]
\caption{Problem with specific labels: results obtained by training the models without segmentation-based cropping or data augmentation.
For each label, the individual models with the best performance and the ensembles that outperform all individual models are marked in bold.
The best ensemble result is marked in italics unless it ties the random classifier.}
\label{table:no_crop_auc_f1}
\resizebox{\textwidth}{!}{
\begin{tabular}{|l|l||ll|ll|ll|ll|ll||ll|ll|ll||} \toprule
                                                &   & \multicolumn{2}{c|}{\bf{DenseNet}} & \multicolumn{2}{c|}{\bf{EfficientNet}} & \multicolumn{2}{c|}{\bf{Inception}} & \multicolumn{2}{c|}{\bf{InceptionResNet}} & \multicolumn{2}{c||}{\bf{Xception}} & \multicolumn{2}{c|}{\bf{PTC-mode}} & \multicolumn{2}{c|}{\bf{PTC-lw}} & \multicolumn{2}{c||}{\bf{CTP}}                        \\ 
                                                              & \# Samples            & AUC              & F1-score  & AUC                & F1-score    & AUC          & F1-score       & AUC             & F1-score          & AUC         & F1-score       & AUC         & F1-score       & AUC        & F1-score      & AUC                     & F1-score \\ \hline 
Normal  & 34327                                                                  & \textbf{0.589}   & 0.470     & 0.500              & 0.374       & 0.500        & 0.374          & 0.500           & 0.374             & 0.500       & 0.374          & 0.500       & 0.374          & 0.500      & 0.374         & 0.589                   & 0.374          \\
Copd signs     & 13419                                                           & 0.500            & 0.457     & 0.500              & 0.457       & 0.500        & 0.457          & 0.500           & 0.457             & 0.500       & 0.457          & 0.500       & 0.457          & 0.500      & 0.457         & 0.500                   & 0.457          \\
Cardiomegaly        & 8412                                                      & \textbf{0.620}   & 0.551     & 0.611              & 0.563       & 0.500        & 0.475          & 0.500           & 0.475             & 0.500       & 0.475          & 0.500       & 0.475          & 0.500      & 0.475         & \textbf{0.633}          & 0.475           \\
Aortic elongation  & 1399           & 0.538            & 0.509     & \textbf{0.553}     & 0.526       & 0.500        & 0.479          & 0.500           & 0.479             & 0.500       & 0.479          & 0.500       & 0.479          & 0.500      & 0.479         & \textbf{0.558}          & 0.479           \\
Unchanged      & 1311                                                           & \textbf{0.535}   & 0.483     & 0.526              & 0.504       & 0.500        & 0.480          & 0.500           & 0.480             & 0.500       & 0.480          & 0.500       & 0.480          & 0.500      & 0.480         & \textbf{0.543}          & 0.480           \\
Scoliosis      & 1073                                                           & 0.500            & 0.484     & \textbf{0.550}     & 0.522       & 0.500        & 0.484          & 0.500           & 0.484             & 0.500       & 0.484          & 0.500       & 0.484          & 0.500      & 0.484         & \textbf{0.550}          & 0.484           \\
Chronic changes    & 873            & \textbf{0.581}   & 0.481     & 0.578              & 0.451       & 0.500        & 0.487          & 0.500           & 0.487             & 0.500       & 0.487          & 0.500       & 0.487          & 0.500      & 0.487         & \textbf{0.585}          & 0.487            \\
Costophrenic angle blunting  & 703  & \textbf{0.556}   & 0.525     & 0.541              & 0.532       & 0.500        & 0.490          & 0.500           & 0.490             & 0.500       & 0.490          & 0.500       & 0.490          & 0.500      & 0.490         & 0.545                   & 0.490            \\
Air trapping        & 663                                                      & 0.500            & 0.490     & 0.498              & 0.510       & 0.500        & 0.490          & 0.500           & 0.490             & 0.500       & 0.490          & 0.500       & 0.490          & 0.500      & 0.490         & 0.498                   & 0.490            \\
Pleural effusion   & 658           & 0.655            & 0.573     & \textbf{0.656}     & 0.567       & 0.500        & 0.490          & 0.500           & 0.490             & 0.500       & 0.490          & 0.500       & 0.490          & 0.500      & 0.490         & \textbf{0.676}          & 0.490            \\
Pneumonia    & 651                                                             & 0.626            & 0.556     & \textbf{0.629}     & 0.566       & 0.500        & 0.490          & 0.500           & 0.490             & 0.500       & 0.490          & 0.500       & 0.490          & 0.500      & 0.490         & \textbf{0.645}          & 0.490            \\
Interstitial pattern  & 594         & \textbf{0.597}   & 0.544     & 0.582              & 0.547       & 0.500        & 0.491          & 0.500           & 0.491             & 0.500       & 0.491          & 0.500       & 0.491          & 0.500      & 0.491         & 0.594                   & 0.491            \\
Infiltrates    & 591                                                           & \textbf{0.615}   & 0.540     & 0.594              & 0.542       & 0.500        & 0.491          & 0.500           & 0.491             & 0.500       & 0.491          & 0.500       & 0.491          & 0.500      & 0.491         & 0.612                   & 0.491            \\
Laminar atelectasis   & 578         & 0.500            & 0.491     & \textbf{0.508}     & 0.491       & 0.500        & 0.491          & 0.500           & 0.491             & 0.500       & 0.491          & 0.500       & 0.491          & 0.500      & 0.491         & 0.508                   & 0.491            \\
Vertebral degenerative  & 575  & 0.500            & 0.491     & \textbf{0.573}     & 0.485       & 0.500        & 0.491          & 0.500           & 0.491             & 0.500       & 0.491          & 0.500       & 0.491          & 0.500      & 0.491         & \textbf{0.573}          & 0.491            \\
Kyphosis     & 526                                                             & \textbf{0.602}   & 0.558     & 0.538              & 0.520       & 0.500        & 0.492          & 0.500           & 0.492             & 0.500       & 0.492          & 0.500       & 0.492          & 0.500      & 0.492         & \textbf{0.606}          & 0.492            \\
Apical pleural thickening & 469    & 0.500            & 0.493     & 0.499              & 0.488       & 0.500        & 0.493          & 0.500           & 0.493             & 0.500       & 0.493          & 0.500       & 0.493          & 0.500      & 0.493         & 0.499                   & 0.493            \\
\begin{tabular}[c]{@{}l@{}}Vascular hilar \\ enlargement\end{tabular}  & 463   & 0.584            & 0.510     & \textbf{0.587}     & 0.475       & 0.500        & 0.493          & 0.500           & 0.493             & 0.500       & 0.493          & 0.500       & 0.493          & 0.500      & 0.493         & \textbf{0.602}          & 0.493            \\
Fibrotic band                         & 449                                    & 0.500            & 0.493     & 0.489              & 0.484       & 0.500        & 0.493          & 0.500           & 0.493             & 0.500       & 0.493          & 0.500       & 0.493          & 0.500      & 0.493         & 0.489                   & 0.493            \\
Nodule      & 449                                                              & 0.500            & 0.493     & 0.500              & 0.493       & 0.500        & 0.493          & 0.500           & 0.493             & 0.500       & 0.493          & 0.500       & 0.493          & 0.500      & 0.493         & 0.500                   & 0.493            \\
\begin{tabular}[c]{@{}l@{}}Calcified \\ granuloma\end{tabular}   & 388         & 0.500            & 0.494     & 0.499              & 0.494       & 0.500        & 0.494          & 0.500           & 0.494             & 0.500       & 0.494          & 0.500       & 0.494          & 0.500      & 0.494         & 0.499                   & 0.494            \\
\begin{tabular}[c]{@{}l@{}}Callus rib \\ fracture\end{tabular}   & 360         & 0.500            & 0.495     & 0.500              & 0.495       & 0.500        & 0.495          & 0.500           & 0.495             & 0.500       & 0.495          & 0.500       & 0.495          & 0.500      & 0.495         & 0.500                   & 0.495            \\
Pacemaker              & 336                                                   & 0.627            & 0.543     & \textbf{0.646}     & 0.523       & 0.500        & 0.495          & 0.500           & 0.495             & 0.500       & 0.495          & 0.500       & 0.495          & 0.500      & 0.495         & \textit{\textbf{0.663}} & 0.495            \\
\begin{tabular}[c]{@{}l@{}}Aortic \\ atheromatosis\end{tabular}   & 318        & 0.500            & 0.495     & \textbf{0.616}     & 0.457       & 0.500        & 0.495          & 0.500           & 0.495             & 0.500       & 0.495          & 0.500       & 0.495          & 0.500      & 0.495         & \textit{0.616}          & 0.495            \\
Volume loss         & 294                                                      & 0.500            & 0.496     & \textbf{0.512}     & 0.496       & 0.500        & 0.496          & 0.500           & 0.496             & 0.500       & 0.496          & 0.500       & 0.496          & 0.500      & 0.496         & \textit{0.512}          & 0.496            \\
Sternotomy             & 292                                                   & 0.530            & 0.517     & \textbf{0.539}     & 0.506       & 0.500        & 0.496          & 0.500           & 0.496             & 0.500       & 0.496          & 0.500       & 0.496          & 0.500      & 0.496         & \textit{\textbf{0.545}} & 0.496            \\
Bronchiectasis              & 290                                              & 0.500            & 0.496     & 0.480              & 0.496       & 0.500        & 0.496          & 0.500           & 0.496             & 0.500       & 0.496          & 0.500       & 0.496          & 0.500      & 0.496         & 0.480                   & 0.496            \\
Hiatal hernia          & 287                                                   & 0.500            & 0.496     & \textbf{0.533}     & 0.506       & 0.500        & 0.496          & 0.500           & 0.496             & 0.500       & 0.496          & 0.500       & 0.496          & 0.500      & 0.496         & \textit{\textbf{0.533}} & 0.496            \\
Pseudonodule          & 275                                                    & 0.500            & 0.496     & 0.498              & 0.500       & 0.500        & 0.496          & 0.500           & 0.496             & 0.500       & 0.496          & 0.500       & 0.496          & 0.500      & 0.496         & 0.498                   & 0.496            \\
\begin{tabular}[c]{@{}l@{}}Hemidiaphragm \\ elevation\end{tabular}   & 254     & 0.515            & 0.496     & \textbf{0.531}     & 0.496       & 0.500        & 0.496          & 0.500           & 0.496             & 0.500       & 0.496          & 0.500       & 0.496          & 0.500      & 0.496         & \textit{\textbf{0.544}} & 0.496            \\
Alveolar pattern       & 248                                                   & \textbf{0.664}   & 0.531     & 0.663              & 0.503       & 0.500        & 0.496          & 0.500           & 0.496             & 0.500       & 0.496          & 0.500       & 0.496          & 0.500      & 0.496         & \textit{\textbf{0.695}} & 0.496            \\
Increased density       & 239                                                  & 0.528            & 0.513     & \textbf{0.536}     & 0.502       & 0.500        & 0.496          & 0.500           & 0.496             & 0.500       & 0.496          & 0.500       & 0.496          & 0.500      & 0.496         & \textit{\textbf{0.547}} & 0.496            \\
\begin{tabular}[c]{@{}l@{}}Vertebral anterior \\ compression\end{tabular} & 214   & 0.546            & 0.510     & \textbf{0.548}     & 0.487       & 0.500        & 0.497          & 0.500           & 0.497             & 0.500       & 0.497          & 0.500       & 0.497          & 0.500      & 0.497         & \textit{\textbf{0.559}} & 0.497            \\
Suture material         & 210                                                  & 0.500            & 0.497     & \textbf{0.542}     & 0.509       & 0.500        & 0.497          & 0.500           & 0.497             & 0.500       & 0.497          & 0.500       & 0.497          & 0.500      & 0.497         & \textit{0.542}          & 0.497            \\
\begin{tabular}[c]{@{}l@{}}Supra aortic \\ elongation\end{tabular}   & 200     & 0.500            & 0.497     & \textbf{0.503}     & 0.497       & 0.500        & 0.497          & 0.500           & 0.497             & 0.500       & 0.497          & 0.500       & 0.497          & 0.500      & 0.497         & \textit{\textbf{0.504}} & 0.497            \\ \midrule
Global        &                                                            & 0.543            & 0.508     & \textbf{0.547}     & 0.502       & 0.500        & 0.488          & 0.500           & 0.488             & 0.500       & 0.488          & 0.500       & 0.488          & 0.500      & 0.488         & \textit{\textbf{0.558}} & 0.488     \\  \midrule         
\end{tabular}}
\end{table}

First, we trained the models with the images without segmentation-based cropping, normalising and data augmentation.
The results obtained, Table \ref{table:no_crop_auc_f1} and \ref{table:no_crophamming}, show that only two individual models have been able to learn, EfficientNet and DenseNet, while the rest of the models were not able to learn and presented a flat training curve with an AUC of 0.5.
As expected, the ensemble does not work correctly, and therefore the preprocessing step is necessary.

\begin{table}[H]
\caption{Problem with specific labels: global results obtained by the individual models and the ensemble without using segmentation-based cropping and data augmentation techniques.}
\label{table:no_crophamming}
\resizebox{\textwidth}{!}{
\begin{tabular}{l|lllll|lll} \toprule
             & DenseNet & EfficientNet & Inception & InceptionResNet & Xception & PTC-mode & PTC-lw & CTP   \\ \midrule
Hamming Loss & 0.067    & 0.107        & 0.046     & 0.046           & 0.046    & 0.046    & 0.046  & 0.046 \\
AUC          & 0.543    & 0.547        & 0.500     & 0.500           & 0.500    & 0.500    & 0.500  & 0.558 \\
F1-score     & 0.508    & 0.502        & 0.488     & 0.488           & 0.488    & 0.488    & 0.488  & 0.488 \\ \midrule
\end{tabular}}
\end{table}

Tables \ref{table:no_augmentation_auc_f1} and \ref{table:no_augmentation_hamming} show the results training with segmentation-based cropping and normalisation but without applying data augmentation techniques.
At first, it is interesting that Inception does not learn, possibly because it is not able to generalise correctly without data augmentation techniques.
InceptionResNet has the best results in most classes, but EfficientNet achieves the best overall result, achieving an AUC of 0.792 while InceptionResNet scores 0.779.
Comparing Table \ref{tab:specific} with these results shows that the application of data augmentation techniques improves the system performance.
If we focus on the results for the different ensembles, we can see that for all labels, the CTP technique performs better than the two PTC methods.
CTP also performs better than the individual models except in three cases: in one case it equals them, and in two cases it performs worse).
We can conclude that data augmentation improves the performance of the system.

\begin{table}[H]
\caption{Problem with specific labels: results obtained by training the models with segmentation-based cropping and normalisation, but without data augmentation. For each label, the individual models with the best performance and the ensembles that outperform all individual models are marked in bold. The best ensemble result is marked in italics.}
\label{table:no_augmentation_auc_f1}

\resizebox{\textwidth}{!}{
\begin{tabular}{|l|l||llllllllll||ll|ll|ll||} \toprule
                                                            &             & \multicolumn{2}{c}{\bf{DenseNet}} & \multicolumn{2}{c}{\bf{EfficientNet}} & \multicolumn{2}{c}{\bf{Inception}} & \multicolumn{2}{c}{\bf{InceptionResnet}} & \multicolumn{2}{c||}{\bf{Xception}} & \multicolumn{2}{c}{\bf{PTC-mode}} & \multicolumn{2}{c|}{\bf{PTC-lw}} & \multicolumn{2}{c||}{\bf{CTP}}                   \\  
                                             & \# Samples        & AUC              & F1-score  & AUC                & F1-score    & AUC          & F1-score       & AUC                 & F1-score      & AUC              & F1-score   & AUC         & F1-score       & AUC         & F1-score      & AUC                     & F1-score  \\ \cline{1-18} 
Normal         & 34327                                                          & 0.5              & 0.374     & 0.802              & 0.725       & 0.453        & 0.374          & \textbf{0.819}      & 0.723         & 0.5              & 0.374      & 0.528       & 0.444          & 0.500         & 0.374         & \textit{0.806}          & 0.374        \\
Copd signs    & 13419                                                           & 0.777            & 0.682     & 0.785              & 0.672       & 0.500          & 0.457          & \textbf{0.799}      & 0.690         & 0.777            & 0.682      & 0.538       & 0.534          & 0.648       & 0.676         & \textit{\textbf{0.825}} & 0.674        \\
Cardiomegaly      & 8412                                                       & 0.900            & 0.768     & 0.898              & 0.774       & 0.641        & 0.474          & \textbf{0.918}      & 0.767         & 0.917            & 0.762      & 0.596       & 0.628          & 0.814       & 0.792         & \textit{\textbf{0.938}} & 0.795         \\
\begin{tabular}[c]{@{}l@{}}Aortic\\ elongation\end{tabular}    & 1399          & 0.863            & 0.700     & 0.874              & 0.686       & 0.500          & 0.479          & \textbf{0.875}      & 0.719         & 0.837            & 0.705      & 0.594       & 0.623          & 0.767       & 0.724         & \textit{\textbf{0.898}} & 0.724         \\
Unchanged      & 1311                                                          & 0.612            & 0.556     & \textbf{0.625}     & 0.549       & 0.500          & 0.480          & 0.597               & 0.549         & 0.602            & 0.544      & 0.506       & 0.495          & 0.531       & 0.539         & \textit{\textbf{0.642}} & 0.537         \\
Scoliosis        & 1073                                                        & 0.823            & 0.678     & 0.808              & 0.690       & 0.500          & 0.484          & \textbf{0.830}      & 0.702         & 0.500            & 0.484      & 0.591       & 0.628          & 0.674       & 0.711         & \textit{\textbf{0.863}} & 0.708         \\
\begin{tabular}[c]{@{}l@{}}Chronic\\ changes\end{tabular}    & 873            & 0.707            & 0.537     & \textbf{0.731}     & 0.553       & 0.500          & 0.487          & 0.696               & 0.547         & 0.695            & 0.538      & 0.515       & 0.518          & 0.625       & 0.568         & \textit{\textbf{0.738}} & 0.568          \\
\begin{tabular}[c]{@{}l@{}}Costophrenic\\ angle blunting\end{tabular} & 703   & 0.810            & 0.698     & 0.837              & 0.691       & 0.500          & 0.489          & \textbf{0.842}      & 0.655         & 0.810            & 0.704      & 0.558       & 0.587          & 0.729       & 0.713         & \textit{\textbf{0.884}} & 0.712          \\
Air trapping      & 663                                                       & 0.500              & 0.490     & 0.671              & 0.568       & 0.500          & 0.490          & 0.500                 & 0.490         & \textbf{0.688}   & 0.553      & 0.508       & 0.506          & 0.500         & 0.490         & \textit{\textbf{0.705}} & 0.490          \\
\begin{tabular}[c]{@{}l@{}}Pleural\\ effusion\end{tabular}    & 658           & 0.925            & 0.839     & 0.942              & 0.818       & 0.479        & 0.046          & \textbf{0.943}      & 0.770         & 0.927            & 0.838      & 0.818       & 0.542          & 0.901       & 0.823         & \textit{0.942}          & 0.825          \\
Pneumonia       & 651                                                         & 0.759            & 0.675     & 0.803              & 0.671       & 0.500          & 0.490          & \textbf{0.808}      & 0.655         & 0.806            & 0.657      & 0.572       & 0.603          & 0.704       & 0.691         & \textit{\textbf{0.851}} & 0.692          \\
\begin{tabular}[c]{@{}l@{}}Interstitial\\ pattern\end{tabular}   & 594        & 0.799            & 0.638     & 0.795              & 0.650       & 0.500          & 0.491          & \textbf{0.813}      & 0.637         & 0.812            & 0.615      & 0.562       & 0.576          & 0.714       & 0.678         & \textit{\textbf{0.858}} & 0.680          \\
Infiltrates     & 591                                                         & 0.733            & 0.620     & 0.776              & 0.635       & 0.500          & 0.491          & \textbf{0.802}      & 0.597         & 0.771            & 0.627      & 0.563       & 0.583          & 0.668       & 0.639         & \textit{\textbf{0.831}} & 0.639          \\
\begin{tabular}[c]{@{}l@{}}Laminar\\ atelectasis\end{tabular}    & 578        & 0.500              & 0.491     & \textbf{0.806}     & 0.639       & 0.500          & 0.491          & 0.754               & 0.630         & 0.745            & 0.646      & 0.560       & 0.587          & 0.572       & 0.607         & \textit{\textbf{0.837}} & 0.607          \\
\begin{tabular}[c]{@{}l@{}}Vertebral\\ degenerative\end{tabular}   & 575      & \textbf{0.730}   & 0.544     & 0.721              & 0.540       & 0.500          & 0.491          & 0.725               & 0.564         & 0.718            & 0.533      & 0.571       & 0.560          & 0.620       & 0.568         & \textit{\textbf{0.771}} & 0.568          \\
Kyphosis     & 526                                                            & 0.796            & 0.611     & \textbf{0.813}     & 0.644       & 0.500          & 0.492          & 0.794               & 0.628         & \textbf{0.813}   & 0.615      & 0.569       & 0.585          & 0.683       & 0.664         & \textit{\textbf{0.860}} & 0.664          \\
\begin{tabular}[c]{@{}l@{}}Apical pleural\\ thickening\end{tabular}  & 469    & \textbf{0.798}   & 0.591     & 0.787              & 0.573       & 0.500          & 0.493          & 0.775               & 0.569         & 0.758            & 0.575      & 0.574       & 0.567          & 0.701       & 0.619         & \textit{\textbf{0.838}} & 0.619          \\
\begin{tabular}[c]{@{}l@{}}Vascular hilar\\ enlargement\end{tabular}  & 463   & 0.679            & 0.562     & \textbf{0.741}     & 0.506       & 0.500          & 0.493          & 0.717               & 0.559         & 0.715            & 0.547      & 0.531       & 0.533          & 0.596       & 0.578         & \textit{\textbf{0.771}} & 0.582          \\
Fibrotic band    & 449                                                        & 0.756            & 0.568     & \textbf{0.772}     & 0.599       & 0.500          & 0.493          & 0.767               & 0.608         & 0.758            & 0.593      & 0.573       & 0.585          & 0.688       & 0.636         & \textit{\textbf{0.813}} & 0.638          \\
Nodule       & 449                                                            & 0.616            & 0.557     & 0.677              & 0.567       & 0.500          & 0.493          & \textbf{0.688}      & 0.547         & 0.626            & 0.558      & 0.535       & 0.545          & 0.566       & 0.574         & \textit{\textbf{0.719}} & 0.572          \\
\begin{tabular}[c]{@{}l@{}}Calcified\\ granuloma\end{tabular}   & 388         & 0.741            & 0.651     & 0.752              & 0.641       & 0.500          & 0.494          & \textbf{0.757}      & 0.622         & 0.689            & 0.611      & 0.578       & 0.601          & 0.645       & 0.654         & \textit{\textbf{0.819}} & 0.656          \\
\begin{tabular}[c]{@{}l@{}}Callus rib\\ fracture\end{tabular}  & 360          & 0.682            & 0.600     & \textbf{0.773}     & 0.594       & 0.500          & 0.495          & 0.500                 & 0.495         & 0.497            & 0.495      & 0.529       & 0.543          & 0.500         & 0.495         & \textit{\textbf{0.799}} & 0.495          \\
Pacemaker         & 336                                                       & \textbf{0.996}   & 0.948     & \textbf{0.996}     & 0.945       & 0.500          & 0.495          & \textbf{0.996}      & 0.946         & \textbf{0.996}   & 0.949      & 0.741       & 0.799          & 0.992       & 0.951         & \textit{0.996}          & 0.951          \\
\begin{tabular}[c]{@{}l@{}}Aortic\\ atheromatosis\end{tabular}   & 318        & \textbf{0.812}   & 0.538     & 0.810              & 0.542       & 0.500          & 0.495          & 0.791               & 0.567         & 0.742            & 0.577      & 0.544       & 0.545          & 0.672       & 0.605         & \textit{\textbf{0.852}} & 0.607          \\
Volume loss          & 294                                                    & 0.855            & 0.687     & 0.862              & 0.717       & 0.500          & 0.496          & \textbf{0.882}      & 0.677         & 0.830            & 0.691      & 0.560       & 0.581          & 0.762       & 0.729         & \textit{\textbf{0.910}} & 0.731          \\
Sternotomy       & 292                                                        & 0.991            & 0.945     & 0.991              & 0.939       & 0.500          & 0.496          & \textbf{0.993}      & 0.872         & \textbf{0.993}   & 0.918      & 0.756       & 0.814          & 0.983       & 0.948         & \textit{\textbf{0.996}} & 0.948          \\
Bronchiectasis            & 290                                               & 0.673            & 0.593     & \textbf{0.726}     & 0.597       & 0.500          & 0.496          & 0.719               & 0.587         & 0.725            & 0.576      & 0.541       & 0.563          & 0.594       & 0.613         & \textit{\textbf{0.784}} & 0.614          \\
Hiatal hernia      & 287                                                      & 0.912            & 0.852     & 0.920              & 0.826       & 0.500          & 0.496          & 0.939               & 0.843         & \textbf{0.945}   & 0.726      & 0.747       & 0.784          & 0.877       & 0.870         & \textit{\textbf{0.962}} & 0.872          \\
Pseudonodule       & 275                                                      & 0.632            & 0.524     & \textbf{0.705}     & 0.547       & 0.500          & 0.496          & 0.536               & 0.514         & 0.639            & 0.540      & 0.530       & 0.536          & 0.540       & 0.544         & \textit{\textbf{0.718}} & 0.545          \\
\begin{tabular}[c]{@{}l@{}}Hemidiaphragm\\ elevation\end{tabular}   & 254     & 0.902            & 0.706     & 0.879              & 0.697       & 0.500          & 0.496          & \textbf{0.911}      & 0.696         & 0.891            & 0.687      & 0.749       & 0.709          & 0.816       & 0.751         & \textit{\textbf{0.951}} & 0.751          \\
Alveolar pattern          & 248                                               & 0.791            & 0.626     & 0.834              & 0.603       & 0.500          & 0.496          & \textbf{0.853}      & 0.580         & 0.810            & 0.604      & 0.568       & 0.579          & 0.715       & 0.621         & \textit{\textbf{0.887}} & 0.622          \\
Increased density      & 239                     & 0.580            & 0.551     & 0.586              & 0.521       & 0.500          & 0.496          & \textbf{0.619}      & 0.521         & 0.569            & 0.526      & 0.501       & 0.500          & 0.533       & 0.537         & \textit{\textbf{0.634}} & 0.539          \\
\begin{tabular}[c]{@{}l@{}}Vertebral anterior\\ compression\end{tabular} & 214  & 0.640            & 0.536     & \textbf{0.645}     & 0.530       & 0.500          & 0.497          & 0.644               & 0.537         & 0.623            & 0.517      & 0.516       & 0.522          & 0.524       & 0.524         & \textit{\textbf{0.702}} & 0.525          \\
Suture material        & 210                           & 0.798            & 0.663     & 0.791              & 0.649       & 0.500          & 0.497          & \textbf{0.824}      & 0.628         & 0.786            & 0.665      & 0.622       & 0.622          & 0.742       & 0.679         & \textit{\textbf{0.833}} & 0.680          \\
\begin{tabular}[c]{@{}l@{}}Supra aortic\\ elongation\end{tabular}   & 200     & 0.697            & 0.569     & 0.778              & 0.564       & 0.500          & 0.497          & \textbf{0.832}      & 0.561         & 0.738            & 0.554      & 0.579       & 0.574          & 0.613       & 0.577         & \textit{\textbf{0.861}} & 0.578          \\ \cline{1-18} 
Global       &                                                            & 0.751            & 0.633     & \textbf{0.792}     & 0.648       & 0.502        & 0.475          & 0.779               & 0.636         & 0.750            & 0.622      & 0.584       & 0.586          & 0.677       & 0.650         & \textit{\textbf{0.831}} & 0.651               \\ \cline{1-18} 
\end{tabular}}
\end{table}

\begin{table}[H]
\caption{Problem with specific labels: global results obtained by the individual models and the ensemble with preprocessing (segmentation-based cropping and normalisation) but without data augmentation.}
\label{table:no_augmentation_hamming}
\resizebox{\textwidth}{!}{
\begin{tabular}{l|lllll|lll} \toprule
             & Densenet201 & EfficientNet & Inception & InceptionResnet & Xception & PTC-mode & PTC-lw  & CTP  \\ \midrule
Hamming Loss & 0.077       & 0.079        & 0.072     & 0.070           & 0.077    & 0.056    & 0.057  & 0.057   \\
AUC          & 0.751       & \textbf{0.792}        & 0.502     & 0.779           & 0.750    & 0.584  & 0.677  & \textbf{0.831}   \\ 
F1-score     & 0.633       & 0.648        & 0.475     & 0.636           & 0.622    & 0.586    & 0.650  & 0.651   \\ \midrule
\end{tabular}
}
\end{table}

\subsection{Performance analysis of CNN models}

The first step is the comparison of the different architectures explained in Section \ref{subsec.cnn}.
They are used as a baseline to compare the ensemble system.
As explained in Section \ref{sec:Methodology}, we consider two types of classification problems: the first uses the original labels proposed by the authors of the dataset ("specific classes"), and the second uses general classes constructed by grouping specific labels.
In the first problem, a finer-grained classification is performed, but it contains a small number of labels, 35, as many of the original 144 do not pass the filter of the minimum number of samples (200).
In the second problem, general radiological patterns are classified, but there is a larger number of labels, 54, because when grouping labels there are a larger number of classes satisfying the minimum threshold of 200 samples.

Tables \ref{tab:specific} and \ref{tab:specific_hammimg} show the results obtained by applying the proposed methodology for the first case study (classification using specific classes).
The model with the best global AUC value is DenseNet, followed by EfficientNet, with 0.818 and 0.804 respectively. The other models (Inception, InceptionResNet and Xception) do not achieve an AUC=0.8.
These results are broken down by class.
First of all, we can observe that the labels with fewer samples do not show worse results on average than the classes with more samples, which means that we have managed to overcome the data imbalance problems of.
It can also be seen in the table that some models perform better with majority classes, such as Inception; others achieve the best results for minority classes, such as EfficientNet and Xception. However, DenseNet 201 and InceptionResNet perform well in both cases.

\begin{table}[H]
\caption{Problem with specific labels: results obtained with by training the models with segmentation-based cropping normalisation and data augmentation. For each label, the individual models with the best performance and the ensembles that outperform all individual models are marked in bold. The best ensemble result is marked in italics.}
\label{tab:specific}
\resizebox{\textwidth}{!}{
\begin{tabular}{|l|l||ll|ll|ll|ll|ll||ll|ll|ll||} \toprule
                                                                           &  & \multicolumn{2}{c|}{\bf{Densenet201}} & \multicolumn{2}{c|}{\bf{EfficientNet}} & \multicolumn{2}{c|}{\bf{Inception}} & \multicolumn{2}{c|}{\bf{InceptionResnet}} & \multicolumn{2}{c||}{\bf{Xception}} & \multicolumn{2}{c|}{\bf{PTC-mode}} & \multicolumn{2}{c|}{\bf{PTC-lw}}  & \multicolumn{2}{c||}{\bf{CTP}}            \\ 
                                                              & \# Samples             & AUC               & F1-score    & AUC                & F1-score    & AUC              & F1-score   & AUC                 & F1-score      & AUC              & F1-score   & AUC          & F1-score       & AUC            & F1-score & AUC         & F1-score       \\ \hline
Normal & 34327                                                                    & 0.820             & 0.722       & 0.811              & 0.716       & \textbf{0.832}   & 0.727      & 0.820               & 0.709         & 0.827            & 0.732      & 0.725      & 0.731      & 0.725  & 0.731    & \textit{\textbf{0.837}} & 0.730    \\
Copd signs      & 13419                                                           & \textbf{0.823}    & 0.681       & 0.785              & 0.644       & 0.816            & 0.678      & 0.815               & 0.675         & 0.800            & 0.666      & 0.588        & 0.610      & 0.647       & 0.675    & \textit{\textbf{0.833}} & 0.672 \\
Cardiomegaly   & 8412                                                            & \textbf{0.927}    & 0.773       & 0.907              & 0.749       & 0.926            & 0.767      & \textbf{0.927}      & 0.777         & 0.922            & 0.779      & 0.746        & 0.765    & 0.825       & 0.789      & \textit{\textbf{0.937}} & 0.791                  \\
\begin{tabular}[c]{@{}l@{}}Aortic\\ elongation\end{tabular}      & 1399          & 0.885             & 0.690       & 0.846              & 0.655       & 0.882            & 0.676      & \textbf{0.888}      & 0.698         & \textbf{0.888}   & 0.702      & 0.690        & 0.682      & 0.777       & 0.705    & \textit{\textbf{0.894}} & 0.707                  \\
Unchanged        & 1311                                                          & 0.636             & 0.553       & 0.614              & 0.543       & 0.638            & 0.549      & \textbf{0.642}      & 0.544         & 0.636            & 0.547      & 0.531        & 0.537          & 0.545       & 0.551 & \textit{0.641}  & 0.551  \\
Scoliosis       & 1073                                                           & 0.759             & 0.636       & 0.712              & 0.598       & 0.745            & 0.605      & 0.732               & 0.602         & \textbf{0.774}   & 0.661      & 0.630        & 0.637     & 0.671       & 0.659     & \textit{\textbf{0.793}} & 0.664                  \\
\begin{tabular}[c]{@{}l@{}}Chronic\\ changes\end{tabular}    & 873              & 0.759             & 0.518       & 0.720              & 0.549       & \textbf{0.768}   & 0.546      & 0.762               & 0.519         & 0.752            & 0.533      & 0.621        & 0.556      & 0.684       & 0.545    & \textit{\textbf{0.772}} & 0.547               \\
\begin{tabular}[c]{@{}l@{}}Costophrenic\\ angle blunting\end{tabular}   & 703   & \textbf{0.862}    & 0.674       & 0.845              & 0.674       & 0.832            & 0.662      & 0.831               & 0.665         & 0.855            & 0.663      & 0.685        & 0.676      & 0.739       & 0.693    & \textit{\textbf{0.877}} & 0.693    \\
Air trapping     & 663                                                          & \textbf{0.692}    & 0.557       & 0.687              & 0.560       & 0.515            & 0.490      & 0.469               & 0.490         & 0.506            & 0.490      & 0.525        & 0.532       & 0.500         & 0.490   & \textit{\textbf{0.704}} & 0.490                   \\
\begin{tabular}[c]{@{}l@{}}Pleural\\ effusion\end{tabular}    & 658             & \textbf{0.959}    & 0.827       & 0.951              & 0.811       & 0.956            & 0.823      & 0.955               & 0.830         & 0.945            & 0.816      & 0.862        & 0.822     & 0.886       & 0.839     & \textit{\textbf{0.967}} & 0.840                   \\
Pneumonia      & 651                                                            & 0.815             & 0.671       & \textbf{0.821}     & 0.660       & 0.810            & 0.663      & \textbf{0.821}      & 0.672         & \textbf{0.821}   & 0.668      & 0.681        & 0.668      & 0.703       & 0.687    & \textit{\textbf{0.850}} & 0.687               \\
\begin{tabular}[c]{@{}l@{}}Interstitial\\ pattern\end{tabular}    & 594         & 0.834             & 0.625       & 0.828              & 0.636       & \textbf{0.846}   & 0.651      & 0.843               & 0.616         & 0.830            & 0.613      & 0.727        & 0.651     & 0.743       & 0.650     & \textit{\textbf{0.858}} & 0.651    \\
Infiltrates        & 591                                                        & 0.812             & 0.629       & 0.803              & 0.617       & 0.803            & 0.626      & \textbf{0.815}      & 0.633         & 0.808            & 0.639      & 0.649        & 0.635      & 0.662       & 0.644    & \textit{\textbf{0.840}} & 0.646                   \\
\begin{tabular}[c]{@{}l@{}}Laminar\\ atelectasis\end{tabular}   & 578           & \textbf{0.843}    & 0.670       & 0.812              & 0.637       & 0.827            & 0.643      & 0.827               & 0.654         & 0.833            & 0.654      & 0.666        & 0.658    & 0.690       & 0.677      & \textit{\textbf{0.858}} & 0.678 \\
\begin{tabular}[c]{@{}l@{}}Vertebral\\ degenerative\\ changes\end{tabular}  & 575  & 0.779             & 0.545       & 0.730              & 0.544       & 0.774            & 0.546      & \textbf{0.785}      & 0.518         & 0.779            & 0.547      & 0.627        & 0.560      & 0.670       & 0.557    & \textit{\textbf{0.797}} & 0.556    \\
Kyphosis                         & 526                                          & \textbf{0.867}    & 0.640       & 0.834              & 0.589       & 0.845            & 0.587      & 0.849               & 0.609         & 0.839            & 0.625      & 0.691        & 0.617     & 0.736       & 0.639     & \textit{\textbf{0.870}} & 0.640     \\
\begin{tabular}[c]{@{}l@{}}Apical pleural\\ thickening\end{tabular}   & 469     & \textbf{0.808}    & 0.553       & 0.801              & 0.568       & 0.789            & 0.573      & 0.509               & 0.493         & 0.500            & 0.493      & 0.592        & 0.574    & 0.661       & 0.619      & \textit{\textbf{0.830}} & 0.621      \\
\begin{tabular}[c]{@{}l@{}}Vascular hilar\\ enlargement\end{tabular}  & 463     & \textbf{0.746}    & 0.549       & 0.742              & 0.568       & 0.769            & 0.522      & 0.755               & 0.544         & 0.745            & 0.515      & 0.618        & 0.556     & 0.651       & 0.562     & \textit{\textbf{0.783}} & 0.563    \\
Fibrotic band         & 449                                                     & \textbf{0.831}    & 0.583       & 0.809              & 0.600       & 0.813            & 0.614      & 0.575               & 0.493         & 0.806            & 0.611      & 0.641        & 0.604     & 0.716       & 0.658     & \textit{\textbf{0.848}} & 0.659    \\
Nodule          & 449                                                           & \textbf{0.706}    & 0.578       & 0.675              & 0.554       & 0.561            & 0.493      & 0.574               & 0.493         & 0.551            & 0.493      & 0.518        & 0.526     & 0.500         & 0.493     & \textit{0.704}          & 0.493    \\
\begin{tabular}[c]{@{}l@{}}Calcified\\ granuloma\end{tabular}    & 388          & \textbf{0.808}    & 0.653       & 0.802              & 0.649       & 0.542            & 0.494      & 0.554               & 0.494         & 0.496            & 0.494      & 0.572        & 0.593    & 0.500         & 0.494      & \textit{\textbf{0.833}} & 0.494    \\
\begin{tabular}[c]{@{}l@{}}Callus rib\\ fracture\end{tabular}      & 360        & 0.717             & 0.606       & \textbf{0.765}     & 0.557       & 0.614            & 0.495      & 0.609               & 0.495         & 0.571            & 0.495      & 0.550        & 0.549     & 0.500         & 0.495     & \textit{\textbf{0.787}} & 0.495    \\
Pacemaker               & 336                                                   & 0.993             & 0.927       & \textbf{0.997}     & 0.942       & 0.996            & 0.919      & 0.996               & 0.931         & 0.984            & 0.926      & 0.984        & 0.930     & 0.993       & 0.946     & \textit{\textbf{0.997}} & 0.946    \\
\begin{tabular}[c]{@{}l@{}}Aortic\\ atheromatosis\end{tabular}     & 318        & 0.856             & 0.521       & 0.847              & 0.516       & 0.862            & 0.550      & 0.852               & 0.541         & \textbf{0.871}   & 0.559      & 0.739        & 0.556     & 0.786       & 0.558     & \textit{\textbf{0.885}} & 0.557    \\
Volume loss      & 294                                                          & \textbf{0.917}    & 0.693       & 0.902              & 0.657       & 0.904            & 0.636      & 0.896               & 0.672         & 0.902            & 0.640      & 0.789        & 0.670     & 0.809       & 0.693     & \textit{\textbf{0.928}} & 0.697    \\
Sternotomy       & 292                                                          & 0.992             & 0.898       & 0.987              & 0.920       & 0.990            & 0.926      & \textbf{0.995}      & 0.936         & 0.992            & 0.865      & 0.961        & 0.912     & 0.984       & 0.941     & \textit{\textbf{0.997}} & 0.941    \\
Bronchiectasis            & 290                                                 & 0.801             & 0.549       & 0.775              & 0.561       & 0.796            & 0.578      & \textbf{0.805}      & 0.562         & 0.794            & 0.573      & 0.682        & 0.580     & 0.690       & 0.587     & \textit{\textbf{0.820}} & 0.588    \\
Hiatal hernia        & 287                                                      & 0.939             & 0.856       & 0.941              & 0.697       & 0.947            & 0.824      & \textbf{0.964}      & 0.801         & 0.947            & 0.851      & 0.871        & 0.786     & 0.876       & 0.867     & \textit{\textbf{0.967}} & 0.867    \\
Pseudonodule       & 275                                                        & 0.612             & 0.496       & \textbf{0.670}     & 0.550       & 0.598            & 0.496      & 0.589               & 0.496         & 0.545            & 0.496      & 0.536        & 0.543     & 0.500         & 0.496     & \textit{\textbf{0.672}} & 0.496    \\
\begin{tabular}[c]{@{}l@{}}Hemidiaphragm\\ elevation\end{tabular}   & 254       & 0.893             & 0.667       & 0.882              & 0.679       & \textbf{0.915}   & 0.670      & 0.894               & 0.702         & 0.898            & 0.684      & 0.759        & 0.692    & 0.784       & 0.725      & \textit{\textbf{0.934}} & 0.724    \\
Alveolar pattern         & 248                                                  & 0.876             & 0.627       & 0.877              & 0.589       & 0.885            & 0.607      & \textbf{0.895}      & 0.606         & 0.871            & 0.594      & 0.748        & 0.612     & 0.774       & 0.623     & \textit{\textbf{0.911}} & 0.622    \\
Increased density       & 239                                                   & 0.643             & 0.541       & 0.641              & 0.509       & 0.651            & 0.534      & 0.633               & 0.526         & \textbf{0.668}   & 0.535      & 0.545        & 0.531     & 0.547       & 0.544     & \textit{\textbf{0.673}} & 0.547    \\
\begin{tabular}[c]{@{}l@{}}Vertebral anterior\\ compression\end{tabular}  & 214  & \textbf{0.749}    & 0.532       & 0.696              & 0.524       & 0.736            & 0.517      & 0.743               & 0.527         & 0.734            & 0.535      & 0.573        & 0.530    & 0.597       & 0.540      & \textit{\textbf{0.752}} & 0.539    \\
Suture material          & 210                                                  & 0.819             & 0.652       & 0.820              & 0.663       & 0.818            & 0.639      & 0.811               & 0.662         & \textbf{0.822}   & 0.612      & 0.759        & 0.663     & 0.768       & 0.685     & \textit{\textbf{0.847}} & 0.684    \\
\begin{tabular}[c]{@{}l@{}}Supra aortic\\ elongation\end{tabular}   & 200       & 0.865             & 0.576       & 0.821              & 0.541       & 0.880            & 0.546      & \textbf{0.882}      & 0.563         & 0.857            & 0.562      & 0.628        & 0.558     & 0.681       & 0.576     & \textit{\textbf{0.894}} & 0.575    \\ \hline
Global                                                                     &  & \textbf{0.818}    & 0.642       & 0.804              & 0.629       & 0.797            & 0.625      & 0.780               & 0.621         & 0.782            & 0.625      & 0.677        & 0.637     & 0.701       & 0.647     & \textit{\textbf{0.840}} & 0.647    \\ \hline
\end{tabular}
}
\end{table}

Secondly, we have analysed the results obtained with the ensemble techniques, using the individual models as baselines. Interestingly, only the CTP technique improves the individual models, as is also the case in Table \ref{table:no_augmentation_auc_f1}.
If we focus on this ensemble technique, we can see that there are two classes, Pleural effusion and pacemaker, where the results of the individual models are not improved.
These two classes have 658 and 336 samples respectively, i.e. they are not majority classes, so one hypothesis would be that the ensemble performs worse in minority classes.
However, the number of labels for which the ensemble does not outperform the individual models is very small compared to the total.
Furthermore, the ensemble achieves an AUC above 0.85 for more than 40\% of the labels, which is higher than expected.
Since we can observe that the ensemble achieves an AUC higher than 0.9 for classes such as hemidiaphragm elevation, hiatal hernia, or sternotomy, all of them with less than 300 samples, we conclude that class imbalance does not affect our system significantly.
Considering that the model is trained for 35 different classes, reaching an imbalance between majority and minority classes of 1:172, we can say that the performance of the system is sufficiently high, considering its characteristics.

\begin{table}[H]
\caption{Problem of specific labels: global results obtained from the individual models and the ensembles.}
\label{tab:specific_hammimg}
\resizebox{\textwidth}{!}{
\begin{tabular}{l|lllll|lll} \toprule
             & Densenet201 & EfficientNet & Inception & InceptionResnet & Xception & PTC-mode &  PTC-lw  & CTP \\ \midrule
Hamming Loss & 0.082       & 0.078        & 0.081     & 0.077           & 0.074    & 0.063  & 0.065  & 0.065  \\
AUC          & 0.818       & 0.804        & 0.797     & 0.780           & 0.782    & 0.677  & 0.701  & 0.840  \\
F1-score     & 0.642       & 0.629        & 0.625     & 0.621           & 0.625    & 0.637  & 0.647  & 0.647  \\ \midrule
\end{tabular}
}
\end{table}

In the second case study used to validate the proposed methodology, we have grouped the different radiological signs into higher level classes that are more general, as shown in the example of fracture types, Figure \ref{fig.treeTerm}. 
After this grouping, the number of labels passing the minimum 200-sample filter rises to 54 (in the case of specific classes only 35 labels passed this threshold).
Therefore, we now train the system with a larger number of labels, which is closer to the reality of health centres.
Regarding the individual models, we can see that the best model is EfficientNet B0 followed by DenseNet, with an AUC of 0.767 and 0.761, respectively.
The rest of the models have a value lower than 0.75.
Regarding the performance per class of each model, we observe that Xception, EfficientNet and DenseNet perform better in majority classes, while Inception and ResNet perform better in minority classes. 

If we look at the results obtained by the ensemble technique, as in the previous case, CTP is the best performer with an AUC of 0.819, which is an improvement of 0.052 over EfficientNet.
There are four classes where the ensemble performs as well as the best individual model, but there is no class where the individual models perform better than the ensemble.
The number of labels where the ensemble achieves an AUC above 0.85 is slightly lower than in the previous case, 37\%, but more than 50\% of the classes have an AUC greater than 0.8.
This is interesting considering the number of classes (54) and their imbalance.
Although the ensemble performs well, it does not perform well for all classes. For example, with the class "Sclerotic bone lesion" it obtains an AUC close to 0.5.

We can observe that in this case the ensemble further improves the individual models as the improvement over the best individual model is now high.
The combination of different architectures avoids overfitting and improves the generalisation capacity in a problem where classification is more difficult due to the specificities of the dataset (high number of classes, multilabel, class imbalance).
These results demonstrate that this methodology works well on highly imbalanced and multilabel datasets.

\begin{table}[H]
\caption{Problem with general labels: results obtained by training the models with segmentation-based cropping normalisation and data augmentation. For each label, the individual models with the best performance and the ensembles that outperform all individual models are marked in bold. The best ensemble result is marked in italics.}
\label{table:generaClasses}

\resizebox{\textwidth}{!}{
\begin{tabular}{|l|l||ll|ll|ll|ll|ll||ll|ll|ll||}
\hline
                                                                           &  & \multicolumn{2}{c|}{\bf{Densenet201}} & \multicolumn{2}{c|}{\bf{EfficientNet}} & \multicolumn{2}{c|}{\bf{Inception}} & \multicolumn{2}{c|}{\bf{InceptionResnet}} & \multicolumn{2}{c||}{\bf{Xception}} & \multicolumn{2}{c|}{\bf{PTC-mode}} & \multicolumn{2}{c|}{\bf{PTC-lw}}           & \multicolumn{2}{c||}{\bf{CTP}}  \\ 
                                                                     & \# Samples      & AUC               & F1-score    & AUC                & F1-score    & AUC              & F1-score   & AUC                 & F1-score      & AUC              & F1-score   & AUC          & F1-score       & AUC                     & F1-score & AUC         & F1-score       \\ \hline
Normal   & 34327                                                                  & 0.735             & 0.685       & 0.707              & 0.652       & \textbf{0.750}   & 0.691      & 0.723               & 0.658         & 0.732            & 0.674      & 0.702        & 0.693    & 0.690       & 0.691      & \textit{\textbf{0.770}} & 0.691                 \\
Copd signs     & 13419                                                            & 0.771             & 0.629       & 0.761              & 0.649       & 0.771            & 0.618      & \textbf{0.793}      & 0.665         & 0.779            & 0.645      & 0.596        & 0.621          & 0.615       & 0.645  & \textit{\textbf{0.816}} & 0.640                   \\

Cardiomegaly     & 8120                                                          & 0.899             & 0.736       & \textbf{0.904}     & 0.746       & 0.890            & 0.723      & 0.892               & 0.751         & 0.898            & 0.741      & 0.766        & 0.747      & 0.816       & 0.760    & \textit{\textbf{0.923}} & 0.762                    \\
\begin{tabular}[c]{@{}l@{}}Thoracic cage\\ deformation\end{tabular}   & 7778     & 0.706             & 0.603       & \textbf{0.728}     & 0.627       & 0.500              & 0.478      & 0.675               & 0.595         & 0.708            & 0.609      & 0.577        & 0.586     & 0.601       & 0.612     & \textit{\textbf{0.745}} & 0.614                    \\
\begin{tabular}[c]{@{}l@{}}Aortic\\ elongation\end{tabular}    & 7436            & 0.858             & 0.691       & 0.853              & 0.690       & 0.842            & 0.661      & \textbf{0.866}      & 0.683         & \textbf{0.866}   & 0.687      & 0.701        & 0.690          & 0.758       & 0.697  & \textit{\textbf{0.886}} & 0.700                    \\
Infiltrates       & 6706                                                         & 0.794             & 0.686       & \textbf{0.802}     & 0.664       & 0.791            & 0.663      & 0.797               & 0.668         & 0.794            & 0.663      & 0.703        & 0.676          & 0.721       & 0.690 & \textit{\textbf{0.827}} & 0.692                    \\
Unchanged    & 6487                                                              & 0.630             & 0.538       & \textbf{0.636}     & 0.552       & 0.618            & 0.543      & 0.633               & 0.545         & 0.631            & 0.552      & 0.536        & 0.543          & 0.536       & 0.546  & \textit{\textbf{0.652}} & 0.547                    \\
\begin{tabular}[c]{@{}l@{}}Chronic\\ changes\end{tabular}    & 4312              & \textbf{0.759}    & 0.548       & 0.754              & 0.542       & 0.752            & 0.525      & 0.734               & 0.520         & 0.740            & 0.522      & 0.656        & 0.567          & 0.685       & 0.543  & \textit{\textbf{0.768}} & 0.545                    \\
Surgery       & 3928                                                             & 0.813             & 0.730       & 0.815              & 0.766       & 0.750            & 0.722      & 0.766               & 0.713         & \textbf{0.829}   & 0.724      & 0.726        & 0.739          & 0.739       & 0.762  & \textit{\textbf{0.845}} & 0.765                    \\
Atelectasis      & 3565                                                          & \textbf{0.798}    & 0.628       & 0.756              & 0.636       & 0.698            & 0.570      & 0.729               & 0.596         & 0.759            & 0.628      & 0.587        & 0.598          & 0.647       & 0.632  & \textit{\textbf{0.804}} & 0.630                    \\
\begin{tabular}[c]{@{}l@{}}Costophrenic\\ angle blunting\end{tabular}   & 3306   & \textbf{0.845}    & 0.655       & 0.807              & 0.638       & 0.758            & 0.604      & 0.784               & 0.638         & 0.828            & 0.652      & 0.660        & 0.634       & 0.700       & 0.656   & \textit{\textbf{0.864}} & 0.658                    \\
Calcified densities    & 3253                                                    & 0.719             & 0.638       & \textbf{0.751}     & 0.639       & 0.500              & 0.491      & 0.500                 & 0.491         & 0.500              & 0.491      & 0.520        & 0.527    & 0.500         & 0.491      & \textit{\textbf{0.764}} & 0.491                    \\
\begin{tabular}[c]{@{}l@{}}Vertebral\\ degenerative\\ changes\end{tabular}  & 3203   & \textbf{0.744}    & 0.502       & 0.726              & 0.528       & 0.676            & 0.497      & 0.733               & 0.487         & 0.730            & 0.512      & 0.643        & 0.532      & 0.664       & 0.514    & \textit{\textbf{0.751}} & 0.514                    \\
Hilar enlargement     & 3162                                                    & \textbf{0.755}    & 0.549       & 0.732              & 0.544       & 0.699            & 0.551      & 0.738               & 0.533         & 0.731            & 0.538      & 0.618        & 0.562     & 0.649       & 0.560     & \textit{\textbf{0.765}} & 0.565                    \\
Pleural thickening     & 3010                                                    & 0.753             & 0.586       & \textbf{0.773}     & 0.585       & 0.737            & 0.572      & 0.743               & 0.525         & 0.763            & 0.562      & 0.651        & 0.587     & 0.671       & 0.586     & \textit{\textbf{0.790}} & 0.587                    \\
\begin{tabular}[c]{@{}l@{}}Mediastinal\\ enlargement\end{tabular}   & 2813       & 0.795             & 0.643       & 0.798              & 0.668       & 0.774            & 0.688      & 0.778               & 0.675         & \textbf{0.822}   & 0.657      & 0.705        & 0.689     & 0.710       & 0.697     & \textit{\textbf{0.841}} & 0.700                    \\
Air trapping        & 2765                                                       & 0.654             & 0.528       & 0.669              & 0.536       & 0.500              & 0.492      & 0.665               & 0.495         & \textbf{0.672}   & 0.536      & 0.520        & 0.523     & 0.602       & 0.544     & \textit{\textbf{0.692}} & 0.546                    \\
Fracture        & 2529                                                           & \textbf{0.749}    & 0.663       & 0.725              & 0.599       & 0.5              & 0.493      & 0.640               & 0.507         & 0.732            & 0.611      & 0.574        & 0.590     & 0.579       & 0.615     & \textit{\textbf{0.792}} & 0.617                    \\
\begin{tabular}[c]{@{}l@{}}Pleural\\ effusion\end{tabular}     & 2436            & \textbf{0.942}    & 0.738       & 0.927              & 0.782       & 0.930            & 0.720      & 0.935               & 0.735         & 0.937            & 0.771      & 0.878        & 0.762     & 0.900       & 0.775     & \textit{\textbf{0.956}} & 0.775                    \\
Granuloma        & 2306                                                          & 0.500               & 0.493       & \textbf{0.777}     & 0.652       & 0.500              & 0.493      & 0.500                 & 0.493         & 0.500              & 0.493      & 0.513        & 0.519          & 0.500         & 0.493  & \textit{\textbf{0.777}} & 0.493                    \\
Nodule         & 1936                                                            & 0.653             & 0.583       & \textbf{0.679}     & 0.571       & 0.617            & 0.542      & 0.643               & 0.547         & 0.622            & 0.575      & 0.560        & 0.569     & 0.558       & 0.575     & \textit{\textbf{0.707}} & 0.573                    \\
Fibrotic band              & 1781                                                & 0.738             & 0.530       & \textbf{0.747}     & 0.531       & 0.712            & 0.522      & 0.500                 & 0.495         & 0.727            & 0.519      & 0.606        & 0.546   & 0.654       & 0.556       & \textit{\textbf{0.770}} & 0.556                    \\
\begin{tabular}[c]{@{}l@{}}Electrical\\ device\end{tabular}    & 1772            & 0.992             & 0.959       & 0.992              & 0.913       & 0.992            & 0.889      & \textbf{0.994}      & 0.871         & 0.992            & 0.929      & 0.990        & 0.935        & 0.992       & 0.942  & \textit{\textbf{0.997}} & 0.942                    \\
Pneumonia       & 1652                                                           & 0.804             & 0.594       & 0.790              & 0.567       & 0.804            & 0.549      & \textbf{0.813}      & 0.577         & 0.799            & 0.599      & 0.712        & 0.602        & 0.728       & 0.606  & \textit{\textbf{0.854}} & 0.607                    \\
\begin{tabular}[c]{@{}l@{}}Aortic\\ atheromatosis\end{tabular}   & 1581          & 0.834             & 0.502       & 0.830              & 0.540       & 0.813            & 0.477      & 0.840               & 0.524         & \textbf{0.843}   & 0.519      & 0.713        & 0.554     & 0.769       & 0.522     & \textit{\textbf{0.866}} & 0.523                    \\
Pseudonodule       & 1451                                                        & 0.693             & 0.561       & \textbf{0.727}     & 0.553       & 0.500              & 0.496      & 0.500                 & 0.496         & 0.708            & 0.562      & 0.557        & 0.555     & 0.576       & 0.589     & \textit{\textbf{0.759}} & 0.593                    \\
Bronchiectasis        & 1430                                                     & 0.795             & 0.544       & 0.776              & 0.571       & 0.789            & 0.548      & \textbf{0.814}      & 0.539         & 0.779            & 0.561      & 0.639        & 0.575        & 0.696       & 0.573  & \textit{\textbf{0.833}} & 0.574                    \\
Hiatal hernia       & 1362                                                       & 0.916             & 0.813       & 0.892              & 0.796       & 0.906            & 0.773      & 0.918               & 0.804         & \textbf{0.927}   & 0.788      & 0.857        & 0.810      & 0.889       & 0.852    & \textit{\textbf{0.959}} & 0.852                    \\
\begin{tabular}[c]{@{}l@{}}Hemidiaphragm\\ elevation\end{tabular}   & 1231       & 0.814             & 0.651       & \textbf{0.841}     & 0.680       & \textbf{0.841}   & 0.596      & 0.823               & 0.649         & 0.811            & 0.645      & 0.733        & 0.670     & 0.746       & 0.683     & \textit{\textbf{0.890}} & 0.683                    \\
\begin{tabular}[c]{@{}l@{}}Increased\\ density\end{tabular}    & 1133            & 0.633             & 0.497       & \textbf{0.640}     & 0.524       & 0.596            & 0.492      & 0.609               & 0.509         & 0.606            & 0.511      & 0.539        & 0.516     & 0.533       & 0.514     & \textit{\textbf{0.661}} & 0.515                    \\
\begin{tabular}[c]{@{}l@{}}Diaphragmatic\\ eventration\end{tabular}   & 757     & 0.500               & 0.498       & \textbf{0.775}     & 0.586       & 0.500              & 0.498      & 0.500                 & 0.498         & 0.500              & 0.498      & 0.525        & 0.534  & 0.500         & 0.498        & \textit{\textbf{0.775}} & 0.498                     \\
Volume loss      & 684                                                          & 0.802             & 0.542       & 0.776              & 0.580       & 0.809            & 0.531      & \textbf{0.814}      & 0.513         & 0.776            & 0.560      & 0.728        & 0.561        & 0.761       & 0.564  & \textit{\textbf{0.865}} & 0.564                     \\
Adenopathy      & 659                                                           & 0.500               & 0.498       & \textbf{0.697}     & 0.538       & 0.500              & 0.498      & 0.548               & 0.520         & 0.583            & 0.543      & 0.500        & 0.498     & 0.521       & 0.528     & \textit{\textbf{0.715}} & 0.528                     \\
\begin{tabular}[c]{@{}l@{}}Bronchovascular\\ markings\end{tabular}   & 602      & 0.712             & 0.570       & 0.738              & 0.537       & \textbf{0.777}   & 0.576      & 0.765               & 0.545         & 0.704            & 0.585      & 0.685        & 0.592     & 0.703       & 0.585     & \textit{\textbf{0.802}} & 0.584                     \\
Mass       & 574                                                                & 0.707             & 0.621       & 0.715              & 0.608       & 0.744            & 0.570      & 0.732               & 0.574         & \textbf{0.746}   & 0.616      & 0.700        & 0.615     & 0.707       & 0.641     & \textit{\textbf{0.806}} & 0.641                     \\
\begin{tabular}[c]{@{}l@{}}Artificial\\ heart valve\end{tabular}    & 562       & 0.969             & 0.658       & 0.953              & 0.730       & 0.975            & 0.696      & \textbf{0.977}      & 0.727         & 0.972            & 0.713      & 0.941        & 0.736        & 0.968       & 0.730  & \textit{\textbf{0.981}} & 0.731                     \\
Catheter     & 545                                                              & 0.871             & 0.740       & 0.874              & 0.721       & 0.866            & 0.673      & \textbf{0.878}      & 0.639         & 0.861            & 0.717      & 0.799        & 0.724        & 0.848       & 0.773  & \textit{\textbf{0.905}} & 0.773                     \\
\begin{tabular}[c]{@{}l@{}}Suboptimal\\ study\end{tabular}    & 544             & 0.743             & 0.524       & 0.693              & 0.510       & \textbf{0.754}   & 0.522      & 0.697               & 0.531         & 0.727            & 0.506      & 0.666        & 0.526     & 0.681       & 0.539     & \textit{\textbf{0.784}} & 0.540                     \\
\begin{tabular}[c]{@{}l@{}}Pulmonary\\ fibrosis\end{tabular}     & 523          & 0.850             & 0.584       & 0.834              & 0.587       & 0.837            & 0.551      & \textbf{0.864}      & 0.577         & 0.862            & 0.565      & 0.795        & 0.577        & 0.810       & 0.591  & \textit{\textbf{0.892}} & 0.591                     \\

\begin{tabular}[c]{@{}l@{}}Heart\\ insufficiency\end{tabular}   & 520           & 0.875             & 0.541       & 0.877              & 0.555       & 0.896            & 0.546      & \textbf{0.884}      & 0.538         & 0.870            & 0.547      & 0.819        & 0.553        & 0.856       & 0.551  & \textit{\textbf{0.920}} & 0.551                     \\
Hypoexpansion             & 476                                                 & 0.838             & 0.541       & 0.745              & 0.545       & \textbf{0.846}   & 0.534      & 0.768               & 0.571         & 0.500              & 0.499      & 0.651        & 0.556      & 0.677       & 0.571    & \textit{\textbf{0.900}} & 0.573                     \\
Gynecomastia         & 437                                                      & 0.852             & 0.527       & 0.810              & 0.552       & 0.852            & 0.501      & \textbf{0.858}      & 0.507         & 0.806            & 0.550      & 0.772        & 0.540        & 0.825       & 0.554  & \textit{\textbf{0.917}} & 0.555                     \\
Emphysema        & 410                                                          & 0.780             & 0.508       & 0.715              & 0.520       & 0.801            & 0.512      & \textbf{0.809}      & 0.506         & 0.724            & 0.521      & 0.684        & 0.529        & 0.732       & 0.524  & \textit{\textbf{0.862}} & 0.525                     \\
\begin{tabular}[c]{@{}l@{}}Sclerotic\\ bone lesion\end{tabular}   & 352         & \textbf{0.506}    & 0.511       & 0.500                & 0.499       & 0.500              & 0.499      & 0.500                 & 0.499         & 0.500              & 0.499      & 0.500        & 0.499    & 0.500         & 0.499      & \textit{\textbf{0.506}} & 0.499                     \\
\begin{tabular}[c]{@{}l@{}}Fissure\\ thickening\end{tabular}     & 336          & 0.816             & 0.533       & 0.802              & 0.573       & 0.819            & 0.518      & \textbf{0.842}      & 0.526         & 0.798            & 0.539      & 0.746        & 0.547        & 0.806       & 0.557  & \textit{\textbf{0.891}} & 0.558                     \\
\begin{tabular}[c]{@{}l@{}}Hilar\\ congestion\end{tabular}       & 318          & 0.785             & 0.503       & 0.798              & 0.519       & 0.790            & 0.514      & \textbf{0.827}      & 0.519         & 0.808            & 0.520      & 0.734        & 0.526        & 0.756       & 0.523  & \textit{\textbf{0.896}} & 0.522                     \\
Osteopenia                                                       & 318          & 0.659             & 0.508       & 0.688              & 0.507       & 0.659            & 0.483      & 0.695               & 0.466         & \textbf{0.701}   & 0.497      & 0.611        & 0.500      & 0.647       & 0.500    & \textit{\textbf{0.752}} & 0.500                     \\
Tuberculosis                                                     & 299          & 0.852             & 0.534       & 0.861              & 0.567       & \textbf{0.869}   & 0.561      & 0.824               & 0.559         & 0.805            & 0.597      & 0.760        & 0.577     & 0.848       & 0.592     & \textit{\textbf{0.909}} & 0.592                     \\
Bullas                                                           & 290          & \textbf{0.746}    & 0.520       & 0.685              & 0.532       & 0.739            & 0.524      & 0.715               & 0.512         & 0.651            & 0.549      & 0.667        & 0.543      & 0.714       & 0.547    & \textit{\textbf{0.777}} & 0.547                     \\
\begin{tabular}[c]{@{}l@{}}Hyperinflated\\ lung\end{tabular}     & 272          & 0.715             & 0.506       & 0.630              & 0.502       & \textbf{0.719}   & 0.504      & 0.645               & 0.485         & 0.659            & 0.501      & 0.649        & 0.513     & 0.658       & 0.512     & \textit{\textbf{0.728}} & 0.512                     \\
Cavitation                                                       & 243          & 0.780             & 0.556       & 0.834              & 0.575       & \textbf{0.856}   & 0.546      & 0.789               & 0.539         & 0.823            & 0.590      & 0.679        & 0.555      & 0.823       & 0.585    & \textit{\textbf{0.934}} & 0.585                     \\
\begin{tabular}[c]{@{}l@{}}Mediastinic\\ lipomatosis\end{tabular}  & 212        & 0.648             & 0.499       & \textbf{0.654}     & 0.551       & 0.5              & 0.499      & 0.500                 & 0.499         & 0.500              & 0.499      & 0.520        & 0.514     & 0.500         & 0.499     & \textit{\textbf{0.681}} & 0.499                     \\
Pneumothorax                                                      & 210         & 0.705             & 0.572       & 0.717              & 0.530       & 0.717            & 0.540      & \textbf{0.721}      & 0.518         & 0.620            & 0.592      & 0.596        & 0.546        & 0.630       & 0.572  & \textit{\textbf{0.847}} & 0.573                     \\
\begin{tabular}[c]{@{}l@{}}Vascular\\ redistribution\end{tabular}  & 204        & \textbf{0.774}    & 0.499       & 0.752              & 0.526       & 0.705            & 0.508      & 0.694               & 0.516         & 0.667            & 0.507      & 0.635        & 0.514      & 0.676       & 0.519    & \textit{\textbf{0.837}} & 0.519                     \\ \hline
Global    &                                                                 & 0.761             & 0.589       & \textbf{0.767}     & 0.600       & 0.732            & 0.566      & 0.739               & 0.572         & 0.739            & 0.589      & 0.669        & 0.594     & 0.696       & 0.601     & \textit{\textbf{0.819}} & 0.602                       \\ \hline
\end{tabular}
}
\end{table}

\begin{table}[H]
\caption{Problem with general labels: global results obtained by the individual models and the ensembles.}
\label{tab:hamming_general}
\resizebox{\textwidth}{!}{
\begin{tabular}{l|lllll|lll} \toprule
             & Densenet201 & EfficientNet & Inception & InceptionResnet & Xception & PTC-mode & PTC-lw & CTP  \\ \midrule
Hamming Loss & 0.070       & 0.065        & 0.070     & 0.075           & 0.065    & 0.052    & 0.057 & 0.056  \\
AUC          & 0.761       & 0.767        & 0.732     & 0.739           & 0.739    & 0.669    & 0.696 & 0.819  \\
F1-score     & 0.589       & 0.600        & 0.566     & 0.572           & 0.589    & 0.594    & 0.601 & 0.602  \\ \midrule
\end{tabular}
}
\end{table}

\subsection{Visual explanation using heatmaps}

\begin{figure}[H]
\centering
    \mbox{\includegraphics[width=6.50in,scale=1]{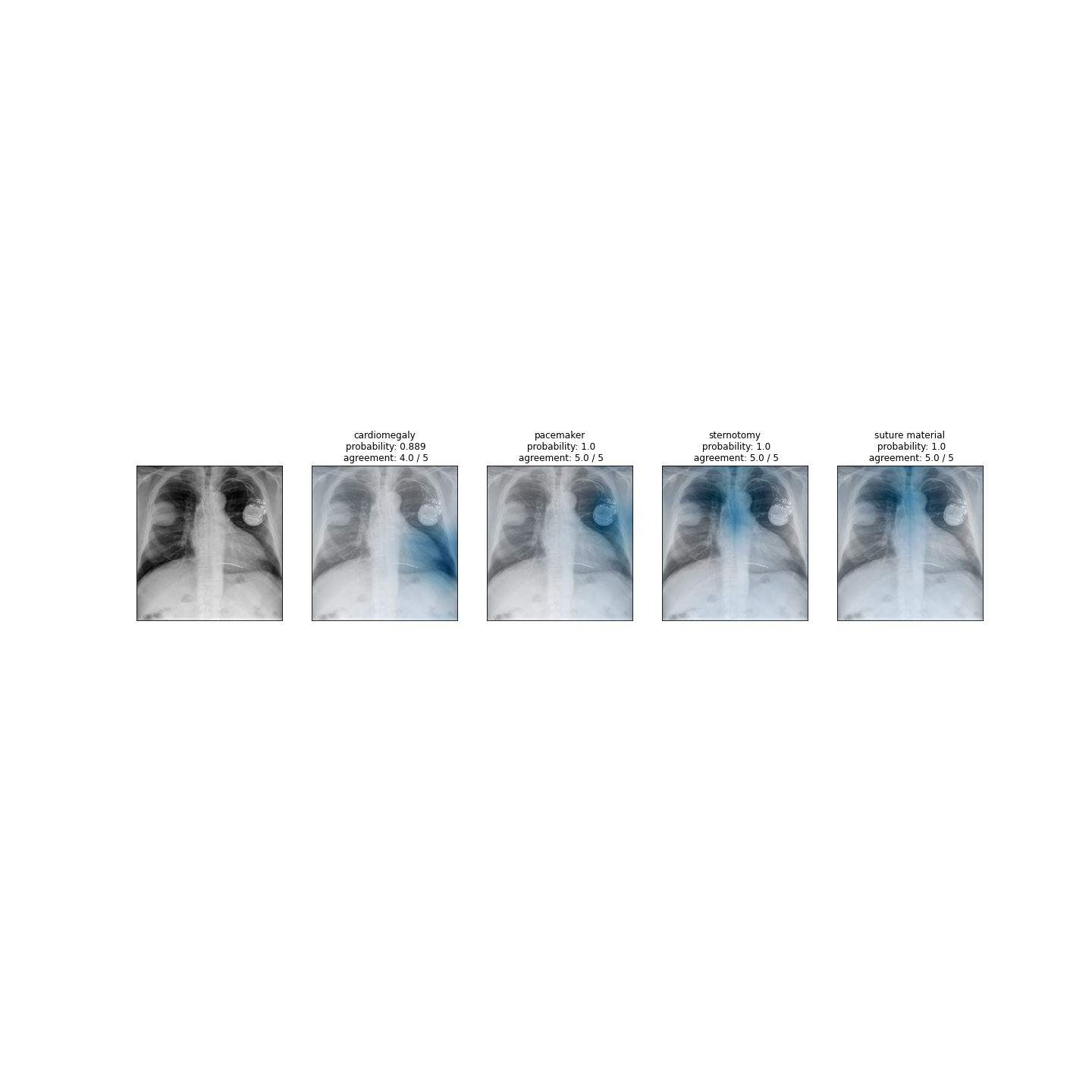}}
  \caption{First visualisation example. The heatmaps of four radiological signs detected (cardiomegaly, pacemaker, sternotomy and suture material) are shown. The title shows the label, the probability estimated by the ensemble, and the agreement between the models of the ensemble. The areas of interest for classification are marked in blue. }
  \label{fig:heatmap1}
\end{figure}

As explained in Section \ref{sec:relatedwork}, the visualisation of multilabel problems is an essential element for this methodology, but it is not a simple problem.
Most of the work in this field has defficiencies.
Therefore, we have developed a technique that for each label generates a heatmap, an estimated probability, and the ensemble agreement.
In Figure \ref{fig:heatmap1}, we can see the original X-ray and the heatmaps of the different classes.
The areas marked on the radiographs match the radiological signs, and the probabilities are high, with three of the four cases showing agreement between all models.

\begin{figure}[H]
  \begin{center}
    \includegraphics[width=\textwidth]{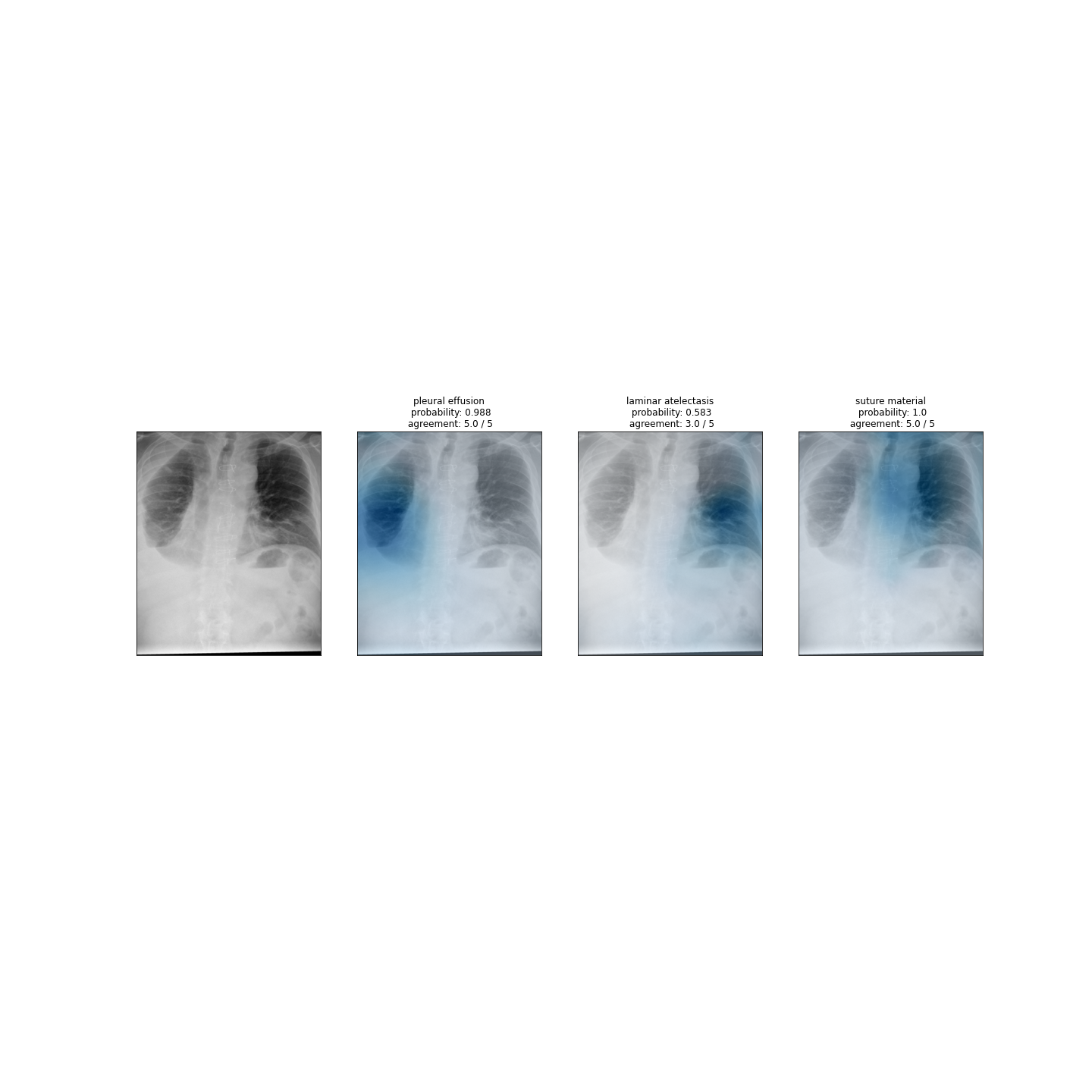}
  \end{center}
  \caption{Second visualisation example. The heatmaps of three radiological signs detected (pleural effusion, laminar atelectasis and suture mataerial) are shown. The title shows the label, the probability estimated by the ensemble, and the agreement between the models of the ensemble. The areas of interest for classification are marked in blue.}
  \label{fig:heatmap2}
\end{figure}

\begin{figure}[H]
  \begin{center}
    \includegraphics[width=0.4\textwidth]{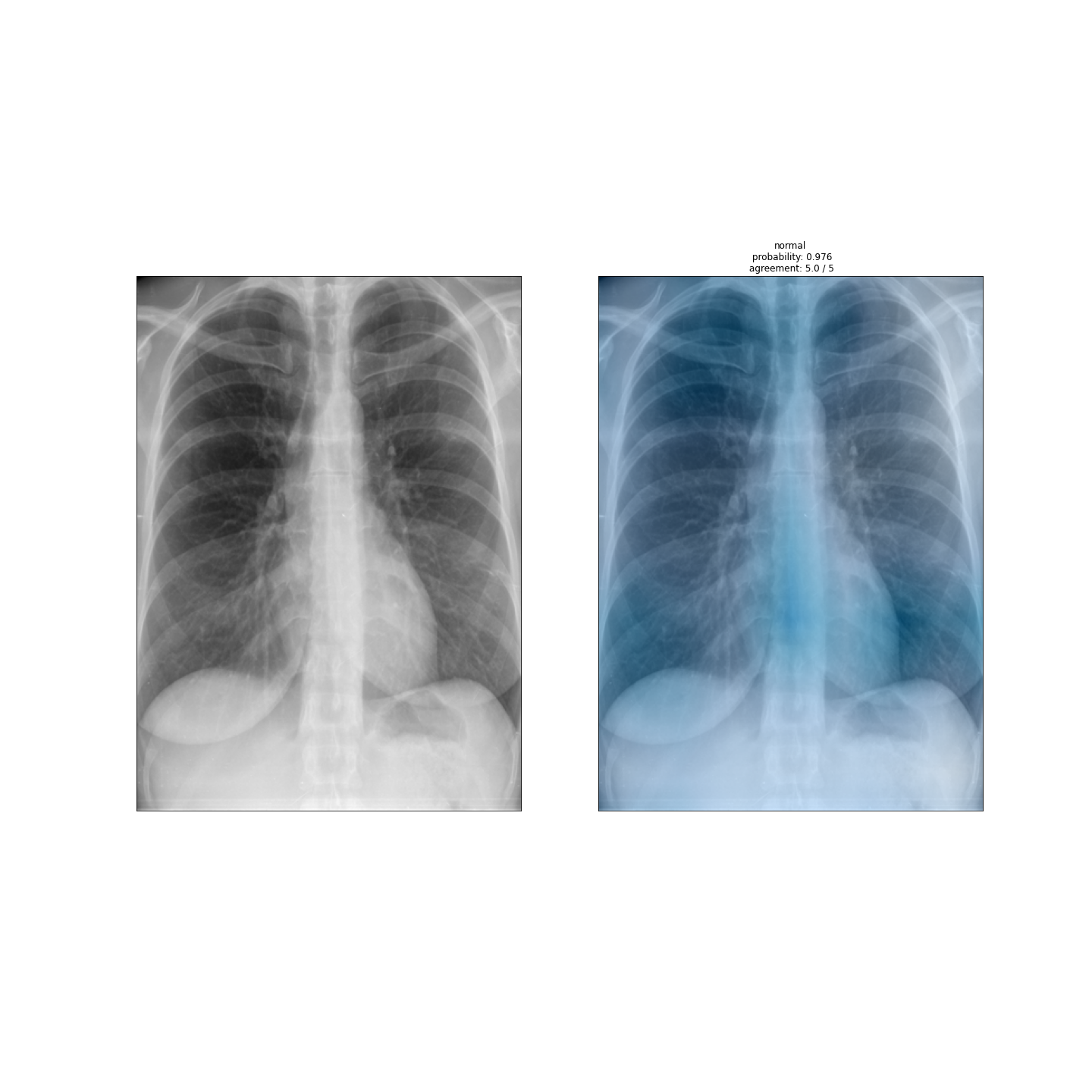}
  \end{center}
  \caption{Third visualisation example. The sample belongs to the normal class. The title shows the label, the probability estimated by the ensemble, and the agreement between the models of the ensemble. The areas of interest for classification are marked in blue.}
  \label{fig:heatmap3}
\end{figure}

In the second example, Figure \ref{fig:heatmap2}, we can see that the class probabilities are lower than before.
The class Atelectasis has an agreement of three models and a low probability (0.583), which means that the physician should be careful with this label.
The last example, Figure \ref{fig:heatmap3}, belongs to the normal class.
In this case, the heat map marks approximately the entire radiograph, as it scans the whole image for radiological signs.
The performance of the visualisations is highly dependent on the performance of the model: if the model is better, the visualisations will be more accurate, and the probability and agreement between models will be higher.
An advantage of this technique over the state of the art is that we generate a grad-CAM map for each sign that includes the probability generated by the system and the agreement between the models of the ensemble.

\section{Discussion}

As mentioned throughout the article, the PadChest dataset has a high quality and is really interesting due to the number of classes, which is higher than other multilabel datasets, and the challenge of class imbalance.
Although we can find numerous papers using this dataset for medical report generation, it is underutilised in chest X-ray classification problems, which makes the available works for comparison scarce.
Moreover, those articles present several problems that complicate an adequate comparison of our work.
Therefore, one of our aims is to generate a methodologically correct baseline that allows comparison for future work. For this purpose, we have conducted two experiments: in the first one we have used the specific radiological signs, i.e. the original ones from the dataset, while in the second we have used more generic radiological signs from a tree of terms provided by the authors of the dataset.
In Table \ref{table:comparativeModels} we can find a summary of the different published systems and their global and class specific performance.
First of all, it is interesting to note how most of the papers have selected different labels to perform the classification, and all papers, except \cite{hashir2020quantifying}, select a low number of total classes compared to the number of classes available. 

\begin{table}[!h]

\centering
\caption{Comparative table of the different state of the art models including their global and class specific performance.}
\label{table:comparativeModels}
\begin{tabular}{|r|ccccc} \toprule
                        & \multicolumn{2}{l|}{Rimeika G. et al. \cite{rimeika2020deep}} & \multicolumn{1}{c|}{Pooch, E. H. \cite{pooch2020can} }         \\ \cline{2-3}
                        & model1            & \multicolumn{1}{c|}{model2}            &  \multicolumn{1}{c|}{}                  \\ \midrule
cardiomegaly            & \multicolumn{1}{c|}{90.36 \%}         & \multicolumn{1}{c|}{91.94 \%}          & \multicolumn{1}{c|}{90.75 \%  }  \\
nodule                  & \multicolumn{1}{c|}{74.97 \%}          & \multicolumn{1}{c|}{71.42 \% }         &  \multicolumn{1}{c|}{    -   } \\
normal                  & \multicolumn{1}{c|}{-}                 & \multicolumn{1}{c|}{-  }               & \multicolumn{1}{c|}{87.10 \%   }  \\
pleural effusion        & \multicolumn{1}{c|}{95.42 \%}          & \multicolumn{1}{c|}{94.93 \% }         & \multicolumn{1}{c|}{-         } \\
pneumonia               & \multicolumn{1}{c|}{-}                 & \multicolumn{1}{c|}{-   }              & \multicolumn{1}{c|}{79.90 \%  } \\
lobar collapse          & \multicolumn{1}{c|}{88.86 \% }         & \multicolumn{1}{c|}{86.39 \%  }        & \multicolumn{1}{c|}{-         } \\
edema                   & \multicolumn{1}{c|}{95.35 \%}          & \multicolumn{1}{c|}{96.05 \%   }       & \multicolumn{1}{c|}{91.07 \%   } \\
subcutaneous emphysema  & \multicolumn{1}{c|}{98.52 \%}          & \multicolumn{1}{c|}{93.79 \% }         & \multicolumn{1}{c|}{    -    } \\
consolidation           & \multicolumn{1}{c|}{87.39 \%}          & \multicolumn{1}{c|}{85.50 \% }         & \multicolumn{1}{c|}{86.07 \%  } \\
pneumothorax            & \multicolumn{1}{c|}{89.95 \%}          & \multicolumn{1}{c|}{88.19 \%  }        & \multicolumn{1}{c|}{82.76 \%  } \\
tuberculosis            & \multicolumn{1}{c|}{92.62 \%}          & \multicolumn{1}{c|}{92.40 \% }         & \multicolumn{1}{c|}{-         } \\
Lymphadenopathy         & \multicolumn{1}{c|}{77.11 \%}          & \multicolumn{1}{c|}{75.81 \% }         & \multicolumn{1}{c|}{-         }  \\
linear atelectasis      & \multicolumn{1}{c|}{84.16 \%}          & \multicolumn{1}{c|}{78.26 \%   }       & \multicolumn{1}{c|}{76.41 \%  } \\
lymph node calficiation & \multicolumn{1}{c|}{82.64 \%}          & \multicolumn{1}{c|}{72.69 \% }         & \multicolumn{1}{c|}{-        } \\
congestion              & \multicolumn{1}{c|}{85.39 \%}         & \multicolumn{1}{c|}{87.29 \% }         & \multicolumn{1}{c|}{-         } \\
Widened mediastinum     & \multicolumn{1}{c|}{75.02 \%}          & \multicolumn{1}{c|}{77.50 \% }         & \multicolumn{1}{c|}{-        } \\
mass                    & \multicolumn{1}{c|}{86.90 \%}         & \multicolumn{1}{c|}{82.29 \% }         & \multicolumn{1}{c|}{   -      } \\
lesion                  & \multicolumn{1}{c|}{- }                & \multicolumn{1}{c|}{- }                & \multicolumn{1}{c|}{69.75 \%  } \\ \midrule

\textit{Global}         & \multicolumn{1}{c|}{\textit{86.98 \%}}    & \multicolumn{1}{c|}{\textit{84.97 \%} }  & \multicolumn{1}{c|}{\textit{82.98 \%} }  \\ \midrule
\end{tabular}
\end{table}

In the case of \cite{rimeika2020deep}, the publication does not show how the two models have been built; it does not provide information on the architecture, the other dataset used, or the criteria for selecting the classes from PadChest dataset, so there is no possibility to replicate these models, and therefore we cannot use it for comparison. In \cite{pooch2020can}, the PadChest classes have been adapted to match the classes of other multilabel datasets such as ChestX-ray14 and CheXpert.
For example, regardless of the fact that the class "Lesion" does not exist in two of the datasets, they generate this class using the ChestX-ray14 labels "Nodules" and "Masses".
However, PadChest was processed in that paper by unifying all classes related to Atelectasis, without providing any medical explanation for this decision, so the labels do not match ours, containing only 8 out of 174 classes available.
Therefore, it cannot be compared with our methodology.
In the case of \cite{hashir2020quantifying}, they first select a single sample from each patient and the authors have used 32 different labels, which is not in line with expectations, since using a lower threshold than ours there should have a larger number of labels.
In this case, we can compare the overall AUC of the system.
These authors achieve an AUC of 0.800 using 32 labels while we obtain 0.8397 using 35 specific labels.
Therefore, we have achieved a better AUC than in \cite{hashir2020quantifying}.
Because of the above reasons, comparing our methodology with the state of the art is really difficult, and therefore one of the aims of this paper is to create a baseline to facilitate the comparison of future work with this dataset.
If we look at the overall AUC of the published models trained with PadChest and compare them with ours, we see that we only outperform two models, but we use a much higher number of classes.
Therefore, we can see that our system performs well and that although it works with a much larger number of labels, it outperforms some of the published models.

\section{Conclusions and future work}
\label{sec:conclusions}

This paper proposes a Deep Learning methodology for classification tasks with imbalanced multilabel datasets.
We have built with this methodology an ensemble of five state-of-the-art architectures: DenseNet-201, EfficientNet B0, Inception, InceptionResNet and Xception.
We have used weighted crossentropy with logit loss to alleviate data imbalance and developed a new technique for generating heatmaps in multilabel classification problems. 

The results of our experiments are promising.
First, in contrast to state-of-the-art papers, we have established a methodologically sound baseline for future work, regardless of whether specific or general classes are used.
It will also allow us to analyse the performance of these models when the number of labels varies.
Our system obtains high AUC values for the number of classes used.
In the case of specific classes, high performance is achieved with an AUC of 0.84.
In the case of general classes, we obtain an AUC of 0.819.
This value may be due to the fact that the general classification has more classes and each of them is composed of different radiological signs.
Thus, the variability is high and it is more difficult to classify.
The results of the visualisation technique show a great potential, as it allows a view of the whole radiograph that differentiates the different pathological signs.
This technique generates a report that includes the visualisation of the heatmap, the probability produced by the system and the agreement between the ensemble models.

There are several ways to improve our methodology.
First, other strategies can be used to alleviate data imbalance, such as adding new samples to the dataset.
This can be done either by obtaining new images from other datasets such as CheXpert, ChestX-ray14, or other single disease datasets, or by creating them with generative adversarial networks (GANs).
Another way to improve system performance is to use different X-ray views of each sample.
To improve the visualisation technique, we can extend the displayed information by including a heatmap that shows the standard deviation of the visualisation of the ensemble \cite{liz2021ensembles}.
This would help medical staff to know in which areas of the heatmap there is more uncertainty.
Another line of work we would like to explore is the generation of a system that returns general and specific labels. In addition to a combination with report generation techniques, doctors would receive a report explaining the different radiological signs and a visual interpretation of these signs.
This could be done using cascade models, which first classify the most general labels and later classify the subcategories.
This would allow to include minority classes, or at least part of them.
Another possible improvement would be to retrain the system using the feedback from experts in the field on the system predictions and heatmaps.

\section*{Acknowledgements}
This research has been supported by the Spanish Ministry of Science and Education under FightDIS (PID2020-117263GB-100) and XAI-Disinfodemics (PLEC2021-007681) grants, by Comunidad Aut\'{o}noma de Madrid under S2018/ TCS-4566 (CYNAMON) and  S2017/BMD-3688 (MULTI-TARGET\&VIEW-CM) grants, by BBVA Foundation grants for scientific research teams SARS-CoV-2 and COVID-19 under the grant: "\textit{CIVIC: Intelligent characterisation of the veracity of the information related to COVID-19}", and by IBERIFIER (Iberian Digital Media Research and Fact-Checking Hub), funded by the European Commission under the call CEF-TC-2020-2, grant number 2020-EU-IA-0252. J. Del Ser thanks the financial support of the Spanish Centro para el Desarrollo Tecnol\'{o}gico Industrial (CDTI, Ministry of Science and Innovation) through the ``Red Cervera'' Programme (AI4ES project), and the support of the Basque Government (3KIA, ref. KK-2020/00049, and the consolidated research group MATHMODE, ref. IT1294-19). Finally, David Camacho has been supported by the Comunidad Aut\'{o}noma de Madrid under "Convenio Plurianual with the Universidad Politécnica de Madrid in the actuation line of \textit{Programa de Excelencia para el Profesorado Universitario}"

\bibliographystyle{unsrt}  

\bibliography{references}

\end{document}